%% file: ms9.tex
\documentclass[twocolumn]{aastex63}

\pdfoutput=1

\usepackage{natbib}
\usepackage{graphicx}
\usepackage{apjfonts}
\usepackage{amsmath}
\usepackage{bm}

\usepackage[ampersand]{easylist}
\ListProperties(Hide=100, Hang=true, Progressive=3ex, Style*=-- )

\hypersetup{urlcolor=cyan}

\newcommand{\kms}       {km~s$^{-1}$}

\newcommand{\pcm}       {cm$^{-2}$}
\newcommand{\lya}       {Ly$\alpha$}

\newcommand{\cgs}       {ergs cm$^{-2}$ s$^{-1}$ \AA$^{-1}$}

\newcommand{\Zemm}    {$Z_\text{H~II}$}
\newcommand{\Zabs}      {$Z_\text{H~I}$}
\newcommand{\Zhost} {$Z_\text{host}$}

\newcommand{\mzs}{$M_* - Z - \text{SFR}$}

\newcommand{\qso}{J0951+3307}
\newcommand{\thisgal}{UGC~5282}

\definecolor{myteal}{RGB}{0, 102, 102}
\definecolor{mydarkgreen}{RGB}{0, 51, 51}

\shorttitle{Absorption Lines from UGC~5282 }
\shortauthors{D. V. Bowen et al}

\received{21 September 2019}
\accepted{9 March 2020}

\begin{document}

\title{ABSORPTION LINE ABUNDANCES IN THE SMC-LIKE GALAXY UGC~5282: 
EVIDENCE OF ISM DILUTION FROM INFLOWS ON KPC SCALES
\footnote{Based on observations with the NASA/ESA {\it Hubble
      Space Telescope} (HST) obtained at the Space Telescope Science
    Institute, which is operated by the Association of Universities for
    Research in Astronomy, Inc., under NASA contract NAS 5-26555.} }

\author{David V.~Bowen}
\affiliation{Princeton University Observatory, Ivy Lane, Princeton, NJ 08544.} 

\author{Todd M. Tripp}
\affiliation{Dept.~of Astronomy, University of Massachusetts,
710 North Pleasant Street, Amherst, MA~01003.}

\author{Edward B.~Jenkins}
\affiliation{Princeton University Observatory, Ivy Lane, Princeton, NJ 08544.} 

\author{Max Pettini}
\affiliation{Institute of Astronomy, University of Cambridge,
Madingley Road, Cambridge, CB3 0EZ, UK.}

\author{Doron Chelouche}
\affiliation{Dept.~of Physics, University of Haifa,
Mount Carmel, Haifa 31905, Israel.}

\author{Renyue Cen}
\affiliation{Princeton University Observatory, Ivy Lane, Princeton, NJ 08544.} 

\author{Donald G.~York}
\affiliation{Dept.\ of Astronomy and Astrophysics, University of
 Chicago,  Enrico Fermi Institute, 5640 South Ellis Avenue, Chicago,
IL~60637.}

\begin{abstract}

  We present an HST {\it Cosmic Origins Spectrograph} (COS) spectrum of the
  QSO SDSS~J095109.12+330745.8 ($z_{\rm{em}} =\, 0.645$) whose sightline
  passes through the SMC-like dwarf galaxy \thisgal\ ($M_B\, =\, -16.0$,
  $cz\, =\, 1577$~\kms ), 1.2~kpc in projection from the central \ion{H}{2}
  region of the galaxy. Damped \lya\ (DLA) absorption is detected at the
  redshift of \thisgal\ with $\log\, [N$(\ion{H}{1})
  cm$^{-2}] \,=\,20.89^{+0.12}_{-0.21}$. Analysis of the accompanying
  \ion{S}{2}, \ion{P}{2} and \ion{O}{1} metal lines yields a neutral gas
  metallicity, \Zabs , of [S/H]$\:\simeq\:$[P/H]$\:=-0.80\pm0.24$.  The
  metallicity of ionized gas from the central \ion{H}{2} region, \Zemm ,
  measured from its emission lines is [O/H]$=\,-0.37\pm0.10$, a difference
  of $+0.43\pm 0.26$ from \Zabs. This difference $\delta$ is consistent
  with that seen towards \ion{H}{2} regions in other star-forming galaxies
  and supports the idea that ionized gas near star forming regions shows
  systematically higher metallicities than exist in the rest of a galaxy's
  neutral interstellar medium (ISM).  The positive values of $\delta$ found
  in \thisgal\ (and the other star forming galaxies) is likely due to
  infalling low metallicity gas from the intergalactic medium that mixes
  with the galaxy's ISM on kpc scales. This model is also consistent with
  broad \lya\ emission detected at the bottom of the DLA absorption, offset
  by $\sim 125$~\kms\ from the absorption velocity. Models of galaxy
  evolution that attempt to replicate population characteristics, such as
  the mass-metallicity relation, may need to start with a galaxy
  metallicity represented by \Zabs\ rather than that measured traditionally
  from \Zemm.
\end{abstract}


\section{INTRODUCTION}
\label{sect_intro}

The history of the universe is essentially the story of how gas, shepherded
by the growth of cold dark matter structures, is turned into stars. The
process is cyclic, with star forming regions inside a galaxy accreting gas
from the intergalactic medium (IGM) and the new stars returning energy and
metals back into the host's interstellar medium (ISM) and the IGM. This
simple ouroboros of inflow and outflow is taken to be a basic ingredient in our
attempts to replicate the universe we see today.

Low mass galaxies provide an important test of our theories about the
growth of galaxies and their evolution. Their star formation history (SFH)
appears to be highly dependent on their mass and their (eventual)
environment \citep[e.g.][and refs.\ therein]{digby18,wright19}, but one way
in which they are different from high mass galaxies is that their shallow
potential wells should allow feedback-driven outflows [from stellar and/or
supernovae (SNae) winds] to significantly impact their ability to retain
metals \citep[e.g.][]{dekel86,MacLow99, ferrara00,garnett02,
  dalcanton07,mcquinn18}.  Star formation can be instigated and, to some
extent, sustained from gas flowing into low-mass galaxies without being
shocked, along 'cold channels' from the IGM \citep[e.g.][]{keres05,dekel06,
  lelli14}, but some hydrodynamical simulations suggest that much of the gas
that is blown out is recycled --- clouds gradually cool, return to the
galaxy \citep{christensen16} and are then re-heated during the next burst
of star formation \citep{muratov17}. As a consequence, the expelled metals
are not quickly reincorporated back into the next generation of stars, and
at $z=0$ the total gas mass of the inner ISM and the outer circumgalactic
medium (CGM) may be similar.

The gas-phase metallicities of the {\it ionized} gas in star-forming dwarf
galaxies can be measured from their \ion{H}{2} emission lines, and they
certainly have some of the lowest abundances known \citep[e.g.][and refs.\
therein]{izotov09,skillman13,hirschauer16}. These abundance estimates can
be used to constrain the fraction of metals that have been retained by a host
\citep[e.g.][]{mcquinn15,gioannini17}. The SFH of dwarfs can be probed
further by examining the {\it ratios} of emission line metallicities: the most
well known example is the change in the nitrogen-to-oxygen ratio (N/O) with
oxygen abundance O/H \citep[e.g.][]{vanZee06,nava06,berg12,james17} as
nitrogen enrichment transitions from a primary to a secondary contribution,
from intermediate-mass stars \citep[e.g.][ands refs.\ therein]{molla06},
pollution by Wolf-Rayet stars \citep{brinchmann08}, or the mixing of
inflowing and outflowing gas \citep{koppen05,amorin10}. A more recent
example is the way in which the variation of the carbon-to-oxygen ratio
with O/H can be understood as the result of 
a series of short bursts of star formation \citep[][and refs.\ therein]{berg19}.

In this paper we measure the abundances in {\it neutral} gas in the nearby
galaxy \thisgal\ using absorption lines detected in the spectrum of the
background QSO SDSS~J095109.12+330745.8 (hereafter ``Q0951+3307'', for
brevity). A comparison between absorption line metallicities in neutral
gas, \Zabs , measured towards background sources, and those measured from
emission lines in ionized gas, \Zemm , from \ion{H}{2} regions within a
galaxy, is important for several reasons.  For sightlines close to
\ion{H}{2} regions, the method offers the opportunity to compare \Zabs\ and
\Zemm\ directly, to test for differences in calibration of \Zemm\ or
whether gas within the \ion{H}{2} regions is more metal rich than the rest
of the galaxy \citep{kunth86}. For QSO sightlines further away from star
forming regions, \Zabs\ can provide a measurement of ISM and CGM metallicity in
areas that cannot be measured in any other way. Such sightlines can probe
gas in the outer regions of a galaxy, where, for example, gas may be
relatively pristine and/or be accreting from the IGM. In addition,
differences in absorption line profiles seen towards \ion{H}{2} regions and
towards outlying QSO sightlines could help constrain the extent of outflowing
gas in a dwarf galaxy's CGM.

Measurements of \Zabs\ towards the {\it same} \ion{H}{2} region used to
measure \Zemm\ have been made using both the {\it Far Ultraviolet
  Spectroscopic Explorer} (FUSE) and the {\it Hubble Space Telescope}
(HST), both of which cover the UV region where suitable absorption lines
lie
\citep{thuan02,aloisi03,etangs04,lebout04,cannon05,thuan05,lebout06,lebout09,lebout13,james14,james18}.
To date, however, a comparison that uses \Zabs\ measured from a background
QSO has only been made once before, towards a probe of the low surface
brightness galaxy SBS~1543+593 \citep{sbs1543_2, regina05}, where \Zabs\
and \Zemm\ were found to be similar. The difficulty, of course, is finding
a QSO bright enough to be observed in the UV with HST, behind a galaxy with
a low enough redshift that emission lines from individual \ion{H}{2}
regions can be recorded. The alignment of \qso\ with \thisgal , only 1.2
kpc from its central \ion{H}{2} region, provides another opportunity to
compare \Zabs\ and \Zemm\ in a low-mass galaxy. These observations were
made as part of an HST program (GO 12486) designed to search for absorption
from several QSO-dwarf galaxy pairs, and this paper presents the first
results from that program.

This paper is organized as follows. \S\ref{sect_galaxy} discusses the
properties of \thisgal, including an image of the galaxy
(\S\ref{sect_image}), an estimation of its star formation rate
(\S\ref{sect_sfr}), and a simple discussion of its environment
\S\ref{sect_environs}. The metallicity of the central \ion{H}{2} region
measured from its emission lines is discussed in \S\ref{sect_h2}. 
The HST COS observations of \qso\ are presented
in \S\ref{sect_cos}, which discusses the damped \lya\ absorption
(\S\ref{sect_dla}) from \thisgal , the corresponding weak metal line
absorption, the resulting absorption line
abundances  (\S\ref{sect_weak}), and several consistency checks of the derived
abundances (\S\ref{sect_check}). The \lya\
emission detected in the  damped \lya\ absorption trough is presented in
\S\ref{sect_lyaemm}. 
The paper
concludes in \S\ref{sect_delta} with a discussion of the difference in \Zemm\ and \Zabs\ for
\thisgal, and compares the value to those found towards \ion{H}{2} regions
in other galaxies. We discuss the implications of our results in \S\ref{sect_discussion}.

\section{UGC~5282 and its Environment}
\label{sect_galaxy}


\begin{deluxetable}{llc}
\tablecolumns{3} 
\tablecaption{Parameters for QSO-galaxy pair \label{tab_properties}}
\tablehead{
\colhead{} & \colhead{} & \colhead{Note}
}
\startdata
\multicolumn{3}{c}{Properties of UGC~5282}\\
\hline
RA, Dec (J2000):                         &   09:51:10.03, +33:07:48.5        & 1 \\
Heliocentric velocity $v_\odot$:    &       $1577\pm3$~\kms              & 2 \\
Adopted Distance $D$:               & $22.0\pm 0.2$ Mpc                      & 3 \\
SDSS mag $g$, $M_g$, $g-r$:     &   15.5, $-$16.3, 0.42                  & 4 \\ 
Johnson mag $B$, $M_B$, $L$:    & 15.7, $-16.0$,  0.02$L_*$           & 5    \\
Radius $R_{25}(r)$:                       & 34$''$ $\equiv 3.7$ kpc              &  6 \\ 
$\mu_c(r)$:                                & 21.7 mags arcsec$^{-2}$             & 6 \\
H~I mass  [$\log(M_\odot)$]:              & $\simeq 8.5$               &   7 \\
Star Formation Rate:                   & $0.05-0.1 \:M_\odot$ yr$^{-1}$           &   8 \\
Stellar mass [$\log(M_\odot)$]:     &  $8.5\pm 0.2$                 &  9 \\
Specific SFR [$\log(\text{yr}^{-1})$]:                              & $-9.5 \pm 0.2$  & 8 \\
Halo mass $\log [M_{200} (M_\odot)$]:           & $\approx 10.7$                         & 10 \\
\hline
\multicolumn{3}{c}{Properties of background QSO} \\
\hline
RA, DEC:                                 &      09:51:09.12 +33:07:45.8         & 11 \\
Redshift:                                 &         0.644                                     & 11 \\
Impact parameter $\rho$:         &      11.7$''$ $\equiv \:$1.2 kpc         &  12 \\
\hline
\enddata
\tablecomments{
(1)  the position of both the brightest central H~II region and the SDSS fiber used to measure galaxy's redshift;
(2) velocity $cz$ of the central H~II region measured in 
\href{http://skyserver.sdss.org/dr12/en/tools/explore/Summary.aspx?id=1237664667895398539} {DR12 of the
SDSS}; 
(3) distance derived from the Tully-Fisher $I$-band spiral luminosity-rotation
correlation listed in the  \href{http://edd.ifa.hawaii.edu/}{Cosmicflows-3
  Distances database} by \citet{tully16}. This is 10\% smaller than would
be inferred from the galaxy's velocity relative to the Cosmic Microwave
Background assuming a concordance cosmology; 
(4) SDSS de-reddened petroMag magnitude and color, and an absolute magnitude assuming $D$;
(5) Johnson magnitudes converted from de-reddened SDSS modelMags using the prescription given by Lupton at 
\href{http://classic.sdss.org/dr7/algorithms/sdssUBVRITransform.html}{this URL}. We
assume $M_* = -20.4$ from
\citet{norberg02}; 
(6) results from ellipse fitting to $r$-band APO image;
(7) $M_{\rm{H~I}} = 2.36\times10^5\:D^2$(Mpc) $I_{\rm{21}}$ $M_\odot$, where
$I_{\rm{21}}$ is taken from \href{http://www.cv.nrao.edu/~rfisher/Arecibo/Profiles/U-05282.109700365.html}{
Aricebo scans available at this URL}, although no errors are cited.
(8) this paper --- see \S\ref{sect_sfr};
(9) using $M/L$ ratios for dwarf galaxies given by  \citet{herrmann16b} and assuming
no errors in the SDSS mags;
(10) from the assumed stellar mass using Fig.~1 of \citet{wright19};  
(11) data from \href{http://skyserver.sdss.org/dr12/en/tools/explore/Summary.aspx?id=1237664667895398537}{SDSS DR12};
(12) the distance between the QSO sightline and the central \ion{H}{2} region, which we take to be
close to the center of the galaxy.
}
\end{deluxetable}
  
\begin{figure*}
\vspace*{0cm}\hspace*{0cm}\includegraphics[width=\textwidth]{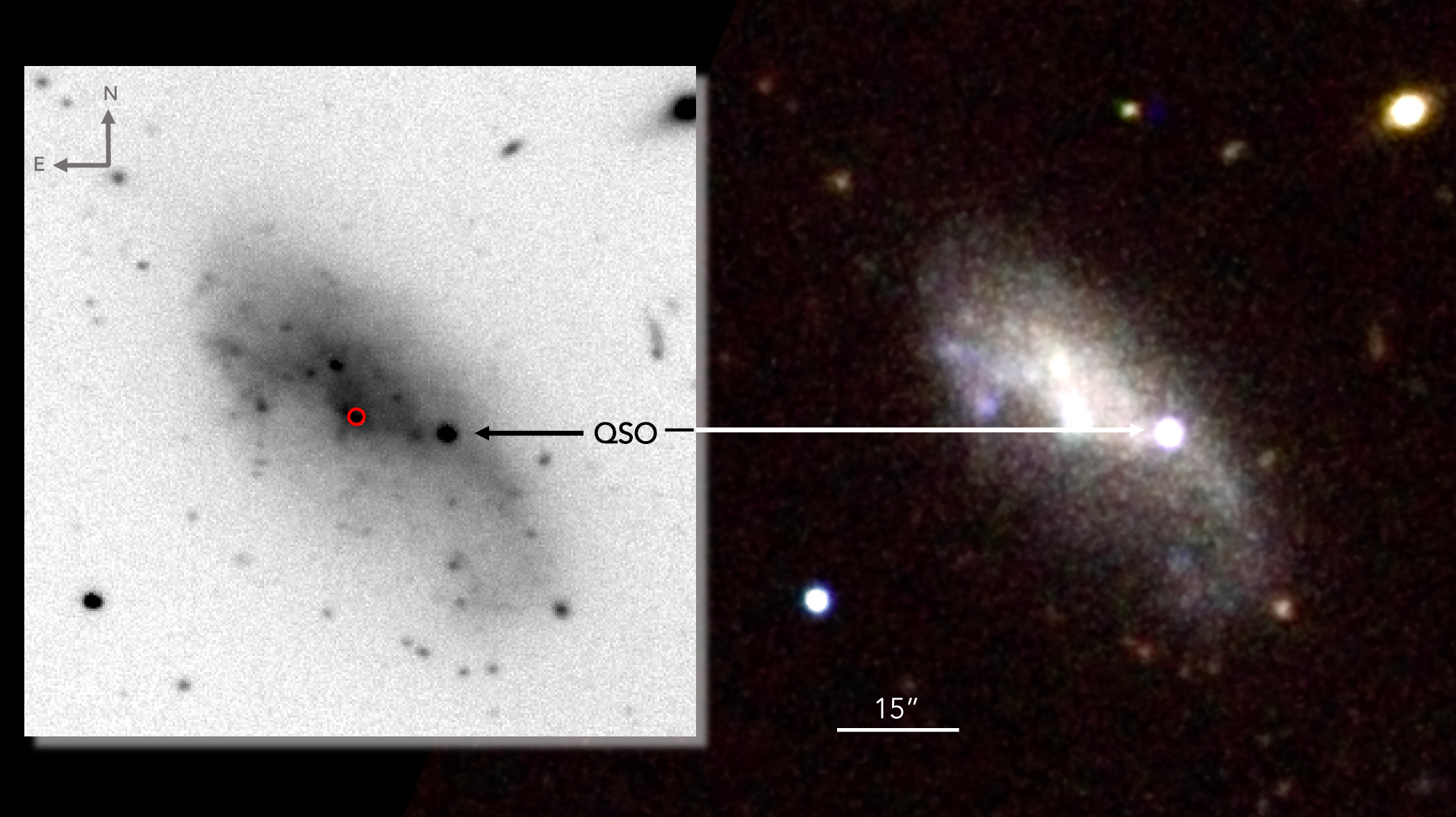}
\caption{{\it Left, inset:} $r-$band image of \thisgal\ taken at APO. Shown in red
  is a 1$''$ radius circle corresponding to the size of the SDSS fiber used
  to measure the redshift of the galaxy, placed at the
  cataloged position of the fiber. We take this to be the center of the
  galaxy. {\it Right:} False color image of the galaxy using SDSS $g-$,
  $r-$ and $i-$band data. These filters were selected because the galaxy is
  barely detected in the $u-$ and $z-$bands. In this color image, the data
  have been smoothed and scaled specifically to highlight structure in the
  galaxy. The 15$"$ scale shown at the bottom of the figure corresponds to
  1.6 kpc assuming a distance of 22~Mpc to the galaxy.
\label{fig_2d}}
\end{figure*}

\subsection{Galaxy Properties}

A collation of some of the properties of \thisgal\ and \qso\ is given in
Table~\ref{tab_properties}.  For comparison, the stellar mass and the
\ion{H}{1} gas mass of the Small Magellanic Cloud (SMC) are both
$\log [M (M_\odot)] \:\simeq\: 8.7$ \citep{mcConnachie12}. Hence \thisgal\
is quite similar to the SMC.  The association of the QSO and the galaxy was
found through a cross-correlation of QSOs discovered by the {\it Sloan
  Digital Sky Survey} (SDSS) as part of Data Release 5 (DR5), with faint
galaxies cataloged by SDSS that had no redshifts.  We obtained a spectrum
of \thisgal\ using the {\it Dual Imaging Spectrograph} (DIS) at the 3.5~m
{\it Apache Point Observatory} (APO) telescope on 2007$-$03$-$18, and found a
redshift from emission lines of $cz=1550\pm10$~\kms . Subsequently, a
higher signal-to-noise (S/N) spectrum was obtained of the central
\ion{H}{2} region as part of SDSS DR12, with $cz = 1577 \pm 3$~\kms .  In
fact, unbeknownst to us at the start of our redshift identification
program, the redshift of the galaxy was first determined from single dish
21~cm observations by \citet{schneider90}, who found $cz = 1557\pm6$~\kms.

\subsection{Galaxy Imaging \label{sect_image}}

We observed \thisgal\ in the $r-$band with the {\it Seaver Prototype
  Imaging camera} (SPIcam) at APO on 11-March-2011 for 1200 sec. The data
were reduced in the normal way for CCD data frames and coadded to produce
the image shown in Figure~\ref{fig_2d}. Conditions were not photometric,
and the final zero-point for the photometry was obtained by tying
magnitudes of objects recorded in our data with their cataloged SDSS
magnitudes.

The $r-$band data from APO is deeper than the $r-$band data recorded by the
SDSS, but the latter also covers $g-$ and $i-$band fluxes which are
recorded with similar S/N. Images from these three bands were smoothed and
aligned to produce the false color image also shown in
Figure~\ref{fig_2d}. The colors are selected to highlight interesting
features in the galaxy, and are not an accurate color representation.

Both images show that \thisgal\ is morphologically irregular, though there
is some indication of a disturbed disk-like structure in the stellar
distribution.  Simulations of dwarf galaxies show that gas disks start to
appear at stellar masses similar to \thisgal\ \citep[e.g.][]{el-badry18},
so the existence of a disc for this galaxy is consistent with such models.
The bulk of the emission is in the north-east half of the galaxy, which
contains several knots of emission from \ion{H}{2} regions, two of which
are very blue. We use the brightest \ion{H}{2} region, observed by SDSS, to
define the center of the galaxy. Fainter \ion{H}{2} regions exist in the
south-west of the galaxy, but the flux from the south-east quadrant is
noticeably less than from the rest of the galaxy.

An $r$-band surface brightness profile for \thisgal\ was constructed from
the APO image using the {\tt ISOPHOTE} package \citep{jedrzejewski87,
  milvang-jensen99} in version 2.1.6 of PyRAF. The fitted ellipses were
constrained to have the same center, i.e. the central \ion{H}{2} region.
 Extrapolating the profile to the center of the galaxy
gives the central surface brightness $\mu_c(r)$ listed in
Table~\ref{tab_properties}.

\subsection{Star Formation Rate of \thisgal \label{sect_sfr}}

\begin{figure}
\begin{center}
\vspace*{0cm}\hspace*{0cm}\includegraphics[width=8.4cm]{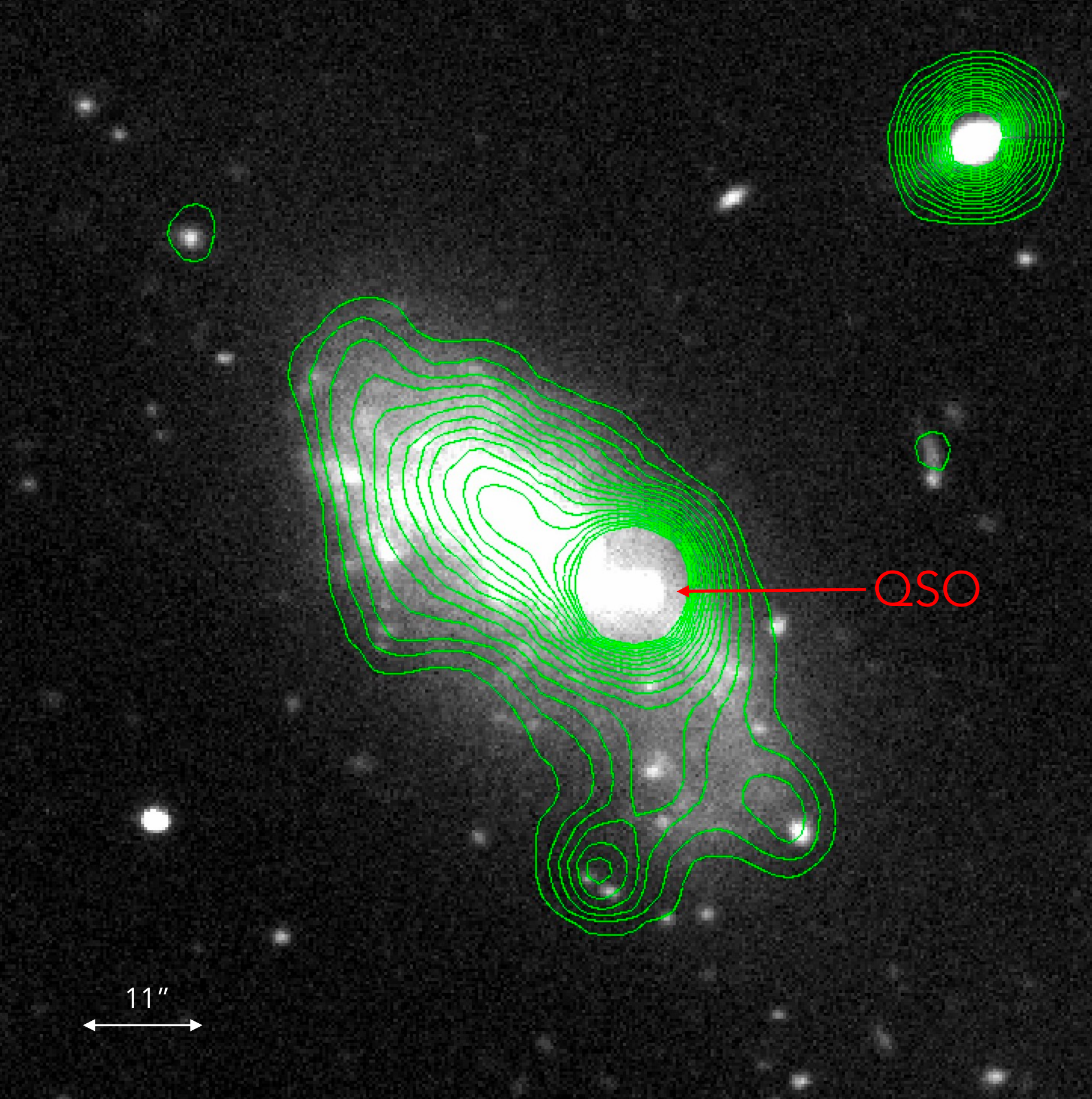}
\caption{The APO $r$-band image shown in Fig.~\ref{fig_2d} with 3.4~$\mu$m
  contours from WISE data superimposed. The contours are linear, and range
  from $1.0-5.9$~$\mu$Jy.
\label{fig_wise1}}
\end{center}
\end{figure}

We estimate the star formation rate (SFR) of \thisgal\ based
on two different methods. The first utilizes {\it Galaxy Evolution
  Explorer} \citep[GALEX, ][]{martin05} images.  We retrieved both the FUV
and NUV data from the GALEX Archive and measured, with the QSO masked out,
magnitudes $m$(FUV)$\:=\:17.5\pm0.1$ and $m$(NUV)$\:=\:17.1\pm0.1$ for the
entire galaxy. We corrected these magnitudes for Milky Way extinction
assuming an extinction $A_{\rm{FUV}}=8.1 \: E$($B-V$) mag
\citep{cardelli89} and a reddening of $E$($B-V$)$\:=0.01$
\citep{schlegel98}. Correcting for extinction by dust in \thisgal\ itself,
however, is more difficult. The total infrared-to-UV flux is often an
indicator of UV extinction as it measures the total stellar emission that
has been absorbed and then re-radiated by dust, relative to the UV light
observed from stars directly. The only infrared (IR) data that exists for
\thisgal\ comes from the {\it Wide-Field Infrared Survey Explorer}
\citep[WISE,][]{wright10}. The galaxy is detected in Band$-1$
(Fig.~\ref{fig_wise1}) and Band$-2$ (3.4 \& 4.6~$\mu$m, respectively), but
not at the longer wavelengths (12 \& 22~$\mu$m); in addition, the QSO is
much brighter than the galaxy at 4.6~$\mu$m, making a measure of the IR
flux from the galaxy unreliable.  For these reasons, we decided not to use
the WISE data to measure the reddening using the IR-to-UV flux ratio.

Instead, we first use the reddening towards the central \ion{H}{2} region.
As we show in \S\ref{sect_h2}, the spectrum supplied by SDSS can be used to
calculate the reddening along the sight line to the star cluster using the
ratio of the Balmer lines, which we found gives an extinction
$A$(H$\alpha$)$\:=0.28\pm0.08$ mag. If we assume that this extinction is
approximately global, and not confined to the sightline to the \ion{H}{2}
region, we can convert $A$(H$\alpha$) to $A$(FUV) by scaling the former by
a factor of 5.8, appropriate for the SMC extinction curve (see
\S\ref{sect_h2}). This gives $A$(FUV)$\:\simeq1.7\pm0.5$ mag, leading to a
corrected FUV magnitude of $15.8\pm0.5$ mags.  The SFR in
$M_\odot$~yr$^{-1}$ is then given by \citet{calzetti13}:

\begin{equation}
\text{SFR(FUV)} = 9\times 10^{-29} \:
L_\nu\text{(FUV)\ (erg s}^{-1} \text{Hz}^{-1} \text{)} 
\label{eqn_kenn98_1}
\end{equation}

\noindent or a SFR of $0.1\pm 0.04 \:M_\odot$~yr$^{-1}$.

Our second estimate of the SFR of \thisgal\ comes from observations made
using a 100~\AA -wide narrow-band H$\alpha$ filter attached to SPIcam at
APO. A total exposure time of 40~mins was spent observing the galaxy
immediately after the $r-$band observations discussed above were
made. After subtracting the $r-$band data to remove the continuum, the
central \ion{H}{2} region could be seen, along with several of the
brightest \ion{H}{2} knots visible in Figure~\ref{fig_2d} and a more
diffuse low surface brightness envelope. As noted above, conditions at APO
during the observations were not photometric, so to calibrate the H$\alpha$
image we matched the counts in a 1$''$ radius aperture placed on the
central \ion{H}{2} region with the H$\alpha$ flux measured by SDSS with the
same sized fiber. After using the same extinction correction $A$(H$\alpha$)
discussed above\footnote{For reference, the values of SFR(FUV) and
  SFR(H$\alpha$) assuming {\bf no} extinction would be $0.02$ and
  0.04$M_\odot$~yr$^{-1}$, respectively.}, we measure the total H$\alpha$
flux within $R_{25}(r)$ to be $9\pm0.1\times 10^{39}$~erg~s$^{-1}$. The SFR
is again given by \citet{calzetti13}:

\begin{equation}
\text{SFR(H$\alpha$)}  = 5.5\times 10^{-42} \:
L\text{(H}\alpha\text{)\ (erg s}^{-1} \text{)} 
\label{eqn_kenn98} 
\end{equation}

\noindent which gives a SFR of $0.05\:M_\odot$~yr$^{-1}$. The uncertainties
in this value are difficult to quantify given the quality of our data: the
background of the $r-$band image was not uniform due to contamination by a
nearby star, and its subtraction from the H$\alpha$ image leads to a
non-uniform background in the latter. In addition, the calibrations of the
SFR for both Equations~\ref{eqn_kenn98_1} and \ref{eqn_kenn98} are
understood to depend on the adopted initial mass function and the
metallicities of the stellar population models.  The difference of a factor
of 2 between our two estimates of the SFR could well be due to the
difficulties in calibrating our narrow-band imaging data. We note, however,
that \citet{lee09} reported that the use of H$\alpha$ tended to
under-predict the SFR compared to values derived from the FUV flux in low
luminosity dwarf galaxies, irrespective of the amount of dust present. If
we use the correction suggested in their Figure~5, we would predict a
SFR(H$\alpha$)$ = 0.06\pm0.03$, which is close to the value we measure.

Contours of the WISE 3.4~$\mu$m emission are shown in
Figure~\ref{fig_wise1}, superimposed on the APO $r$-band data.  Most of the
flux comes from the center of \thisgal , as expected, but there also
appears to be additional IR emission to the south and to the south-west,
where the $r-$band flux is relatively weak. (These regions are also
discernable at 4.6~$\mu$m.) It is possible that these areas have regions of
star formation that are hidden by dust, which would imply that the SFRs
calculated above are only lower limits. A patchy distribution of dust
within the galaxy might also go some way in explaining its irregular
morphology.

\subsection{Galaxy Environment \label{sect_environs}}

\begin{figure*}
\begin{center}
\vspace*{0cm}\hspace*{0cm}\includegraphics[width=18cm]{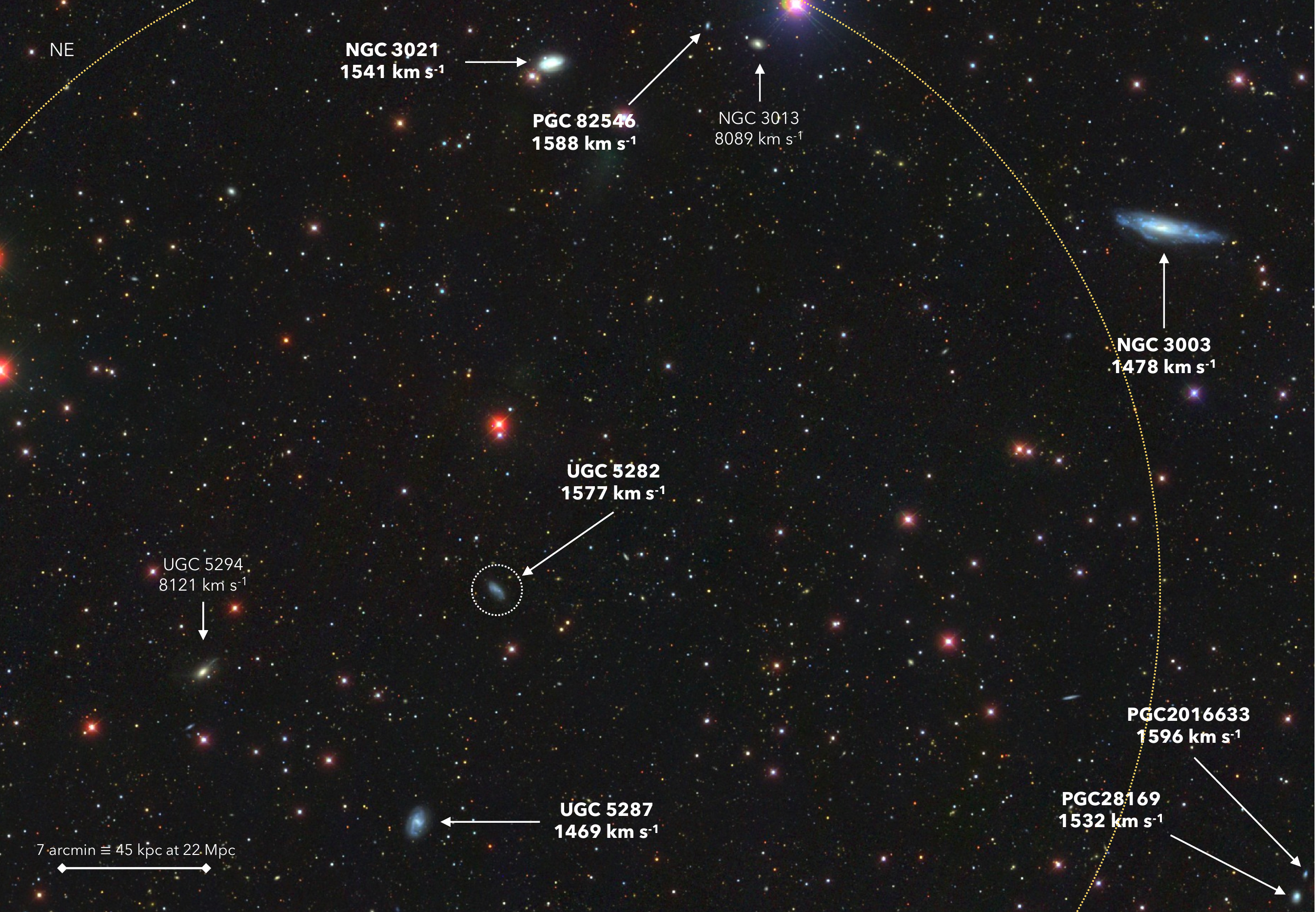}
\caption{SDSS image of the immediate environment around \thisgal. The
  field is dominated by the $M_B = -20.5$ galaxy NGC~3003 which lies
  225~kpc from \thisgal\ on the plane of the sky. Its companion is NGC~3021 ($M_B = -19.6$) which
  lies 198 kpc away. All the galaxies whose redshifts are known and within
  $\pm 400$~\kms\ of the redshift of NGC~3003 are labelled in bold.
The yellow dotted circle represents a
radius of 200 kpc centered on \thisgal\ assuming a distance of 22~Mpc from
the Milky Way.
\label{fig_lss}}
\end{center}
\end{figure*}

The HyperLeda catalog \citep{makarov14} lists 23 galaxies within 1~Mpc and
$\pm 300$~\kms\ of \thisgal, 13 of which are brighter. The dwarf is clearly
part of a galaxy group, labelled by \citet{marino13} as the ``U268'' group
within the Leo cloud \citep{tully88}.  The nearest galaxy to \thisgal\ is
UGC~5287 (a separation of $\rho = 72$~kpc from \thisgal), another blue star
forming dwarf, and a magnitude
brighter  ($M_B = -17.2$) than \thisgal . Both are likely associated with
the bright spiral galaxies NGC~3021 ($\rho = 162$~kpc, $M_B = -19.6$) and
NGC~3003 ($\rho = 225$~kpc, $M_B = -20.5$). All these galaxies are shown in
Figure~\ref{fig_lss}.

In comparison to Local Group galaxies, the configuration of the lower
luminosity galaxies \thisgal\ and UGC~5287 shown in Figure~\ref{fig_lss} is
reminiscent of the LMC and SMC's interactions with our Galaxy, although the
distances between the host (NGC~3003) and its satellites are much larger
than those between the Magellanic Clouds and the Milky Way (MW); given the
values of $M_B$ for both NGC~3021 and NGC~3003, \thisgal\ and UGC~5287
probably lie just beyond the formers' virial radii, and may only be
starting on their passage into the DM halo dominated by NGC~3003. In this
sense, \thisgal\ may be a younger version of the SMC, seen before its more
complex interactions with the MW and the LMC.

For galaxies with masses similar to \thisgal , $\ga 75$~\% are still
forming stars when they lie within 250~kpc of a host galaxy \citep{geha12},
so the presence of a star-forming dwarf galaxy close to the more massive
galaxies is not unusual.  Again, based on comparisons with Local Group
dwarf galaxies and results from simulations, \thisgal , as an irregular
dwarf galaxy, has likely been forming stars for most of its history
\citep{gallart15} and has only recently begun to be accreted by the two
more massive galaxies in the group.  Interactions between \thisgal\ and
UGC~5287 \citep{pearson18,pearson16}, as well as stripping (or partial
stripping) by the halos of NGC~3021 and 3003 \citep[e.g.][and
refs.~therein]{brown17,fillingham16,emerick16,salem15} might
explain the irregular morphology of the galaxy.  Alternatively, the
on-going star formation in \thisgal\ may actually have been triggered by
its interactions with UGC~5287 \citep{stierwalt15, lelli14}. We reiterate
these hypotheses in \S\ref{sect_discussion}.

\section{METALLICITY OF THE CENTRAL H~II REGION}

\label{sect_h2}

As noted above, the central \ion{H}{2} region was observed by SDSS, with
the fiber covering the region shown in Figure~\ref{fig_2d}. The
\href{http://skyserver.sdss.org/dr12/en/tools/explore/Summary.aspx?id=1237664667895398539}{DR12
  SDSS spectrum} (Plate$-$MJD$-$Fiber ID = 5798$-$56326$-$050) shows a
continuum rising slowly towards the blue, superposed with many archetypal
\ion{H}{2} narrow emission lines. These can be used to determine the
\ion{H}{2} region's metallicity.

After correcting the spectrum for extinction by the Milky Way, we modeled
the continuum using the \href{http://www.starlight.ufsc.br/}{ {\tt
    STARLIGHT}} spectral synthesis code described by
\citet{fernandes05}. Not surprisingly, given the lack of any significant
stellar absorption lines, the fluxes of the emission lines were corrected
by only very small amounts. The continuum itself could be modelled with a
young ($\sim 1-3\times10^6$ yr) population of stars that has a total
stellar mass of $1-2\times10^6$~M$_\odot$. After correcting for the
extinction at H$\alpha$ (see below), the H$\alpha$ luminosity of the
\ion{H}{2} region (as covered by the SDSS fiber) is
$\log [L\text{(H}\alpha \text{)(erg s}^{-1}\text{)]} = 38.22\pm0.01$ and
the SFR using Eqn.~\ref{eqn_kenn98} is
$\log [\text{SFR (M}_\odot\:\text{yr}^{-1})] = -3.04\pm0.01$ for this
specific star cluster (and not for the galaxy as a whole).

It is well known that the Balmer emission line fluxes have fixed ratios
with respect to each other under certain assumptions.  For Case-B
recombination, and assuming electron densities and temperatures typical for
an \ion{H}{2} region ($10^2$~cm$^{-3}$ and 10,000~K respectively), ratios
of H$\alpha$/H$\beta = 2.85$, H$\gamma/$H$\beta = 0.469$,
H$\delta/$H$\beta= 0.260$, etc.\ are predicted \citep{hummer87}.
Deviations from these ratios are taken to be due to dust extinction along a
line of sight, with the broadband color excess $E$($B-V$) given by
\citep[e.g.][]{momcheva13}

 \begin{equation}
E(B-V) = \frac{-2.5}{\kappa(\text{H}\alpha)-\kappa(\text{H}\beta)}\times \log_{10} \left[
  \frac{(\text{H}\alpha/\text{H}\beta)_{\text{obs}}}{ (\text{H}\alpha/\text{H}\beta)_{\text{true}}       } \right]
\end{equation}

\noindent where $\kappa$(H$\alpha$) and $\kappa$(H$\beta$) are the values
of a dust attenuation curve at the wavelengths of H$\alpha$ and H$\beta$
and (H$\alpha$/H$\beta$)$_\text{true} = 2.85$. Additional estimates of
$E$($B-V$) can be made if other Balmer lines are present using a similar
relationship. The extinction for any other line at a wavelength $\lambda$
is then simply $A$($\lambda$)$\:=\kappa$($\lambda$)$\:E$($B-V$). We
estimated values of $\kappa$ using the extinction curve given by
\citet{gordon03} for the SMC, with $R_V = 2.74$, where $R_V =
A(V)/E(B-V)$. The measured ratios H$\alpha$, H$\gamma$, and H$\delta$ to
H$\beta$ gave similar values for \thisgal, with H$\alpha$/H$\beta$ having
the smallest errors:

\[
E(B-V) = 0.13 \pm 0.04
\]

\noindent
and from which all the emission lines detected towards the H~II region in
\thisgal\ were  corrected. The extinction at H$\alpha$ (used in the previous
section) was $A$(H$\alpha) = 0.28\pm0.08$ mags.

After correcting the spectrum for this extinction, we measured the emission
line fluxes and their errors.  Fluxes were measured by fitting single
Gaussian profiles (except for [\ion{O}{2}]~$\lambda3727$ where both lines
of the doublet are blended to show a clearly asymmetric profile --- the
flux integrated over the whole line was used instead); errors were
generated by using a Monte-Carlo approach of fitting multiple synthetic
constructions of the initial fitted line, with their errors defined by
those supplied by SDSS. An additional term
$\sigma_c^2 = \sum ( \sigma_i^2)$ was added in quadrature to this error in
order to account for uncertainties in the background, which can be
significant for weak lines, where $\sigma_i$ is the error at the $i$th
pixel given by the error array, summed over the number of pixels used to
define the background.  The spectrum shows Balmer lines down to H$\zeta$
and many forbidden collisionally excited metal lines.  When compared to
values that define the commonly used BPT \citep{baldwin81} relationships,
the ratios of $\log$([\ion{O}{3}]$\lambda 5007/$H$\beta$)$\:=0.20\pm0.02$
and $\log$([\ion{N}{2}]$\lambda 6583/$H$\alpha$)$\:=-1.05 \pm0.03$ place
the source well within the area of star-forming \ion{H}{2} regions.

There are two well-known methods for determining the metallicities of
extragalactic \ion{H}{2} regions from their emission lines; first, there is
the ``direct'', or ``$T_e$'' method, which measures electron temperatures
directly from ratios of weak and strong recombination lines arising from
atomic levels with substantially different excitation levels; second,
strong emission line (SELs) ratios can be used, calibrated by using ether
photoionization models, or by using \ion{H}{2} regions for which O/H has
already been measured using the $T_e$ method.  In addition, it is possible
to use recombination lines (RLs) of heavy elements to RL lines of hydrogen
to measure O/H (the ``RL method''), but the metal lines are often too weak
to be observed in many extragalactic objects.

The resolution and S/N of the SDSS spectrum is too low to permit detection
of weak auroral lines such as [\ion{O}{3}]~$\lambda 4363$ or
[\ion{N}{2}]~$\lambda 5755$, which are often used to calculate $T_e$ in an
\ion{H}{2} region.  We therefore used SEL ratios to measure \Zemm . We
avoided using SEL diagnostics which rely on the [\ion{O}{2}]~$\lambda 3727$
line simply because the correction of fluxes from reddening due to internal
dust extinction are the most severe in the blue.  Fortuitously, the long
wavelength range of the SDSS spectrum covers [\ion{Ar}{3}]~$\lambda 7135$
and [\ion{S}{3}]~$\lambda\lambda 9069, 9530$ lines in the red. Since S, Ar
and O are all $\alpha$ elements and are produced by the same types of
stars, the Ar3O3 and S3O3 indexes calibrated by \citet{stasinska06} provide
a measurement of the oxygen abundance O/H that should be largely unaffected
by chemical evolution effects.  The definitions of both indexes are listed
in Table~\ref{tab_HiiZ}. Another index which also uses both [\ion{S}{2}]
and [\ion{S}{3}] lines is S23 \citep{diaz00,perez-montero05}.
Table~\ref{tab_HiiZ} shows that Ar3O3, S3O3 and S23 all give consistent
results: 12+$\log\:$O/H $ \sim 8.4$, with an error likely dominated by the
errors in the index calibration, between $\simeq \pm 0.1$ to $\pm 0.25$
dex.

\begin{deluxetable*}{llcllc}
\tablecolumns{6} 
\tablecaption{Strong Line Diagnostics of The Central H~II Region of 
  \thisgal \label{tab_HiiZ} }
\tablehead{
\colhead{Method} 
& \colhead{Method Line} 
& \colhead{Method} 
& \colhead{} 
& \colhead{} 
& \colhead{Calibration} \\
 \colhead{ID}
& \colhead{Ratios} 
& \colhead{$\sigma_{\text{SL}}$\tablenotemark{a}}
& \colhead{12 + $\log$(O/H)}
& \colhead{$Z_\text{H II}$\tablenotemark{b}}
&\colhead{Ref.}
}
\startdata
\hline
Ar3O3   & $[$\ion{Ar}{3}$]$~$\lambda 7135$ / $[$\ion{O}{3}$]$~$\lambda 5007$ & $\pm 0.23$     & $8.43\pm 0.10$ & $-0.33 \pm 0.11$ & 1 \\
S3O3     & $[$\ion{S}{3}$]$~$\lambda 9069$ / $[$\ion{O}{3}$]$~$\lambda 5007$ & $\pm 0.25$      & $8.42\pm0.03$  & $-0.34\pm 0.06$  & 1 \\
S23       & ($[$\ion{S}{2}$]$~$\lambda\lambda 6716, 6730$ + $[$\ion{S}{3}$]$~$\lambda\lambda 9069,9530$)/H$\beta$  &   $\pm 0.10$  & $8.39\pm0.03$ & $-0.37 \pm 0.06$ & 2\\
N2        & $[$\ion{N}{2}$]$~$\lambda 6583$ / H$\alpha$                                   & $\pm 0.16$     & $8.26\pm0.01$    & $-0.50 \pm0.05$ & 3 \\
O3N2    & ($[$\ion{O}{3}$]$~$\lambda 5007$ x  H$\alpha$) /
($[$\ion{N}{2}$]$~$\lambda 6583$ x H$\beta$) & $\pm 0.18$ & $8.27\pm0.01$  
              & $-0.49 \pm0.05$ & 3 \\
KK04    & $f$([\ion{O}{2}]~$\lambda 3727$, [\ion{O}{3}]~$\lambda 4959,
5007$, H$\beta$, $q$)    & $\pm 0.2\phn{}$ & 8.36  & $-0.40$  & 4 \\
\hline
\enddata
\tablenotetext{a}{This is the the approximate 1$\sigma$ dispersion in the calibration of the
  emission line ratio metallicities.}
\tablenotetext{b}{Metallicity of the central \ion{H}{2} region in \thisgal , $\log$(O/H) $-\log$(O/H)$_\odot$ where
  $12+\log$(O/H)$_\odot = 8.76$ \citep{lodders03}. The errors listed in this column are only
  combined errors from the flux measurements.}
\tablerefs{1 --- \citet{stasinska06}; 
2 --- \citet{perez-montero05}; 
3 --- \citet{marino13_ra}; 
4 --- \citet{kobulnicky04}}
\end{deluxetable*}

Table~\ref{tab_HiiZ} also lists the [\ion{N}{2}]~$\lambda 6583$ line
diagnostics N2 and O3N2 \citep[e.g.][and refs
therein]{alloin79,storchi-bergmann94, pettini04, marino13_ra}. N2 in
particular is highly sensitive to the O/H abundance and it is largely
unaffected by errors in the reddening correction, because of the similarity
in the wavelengths of [\ion{N}{2}]~$\lambda 6583$ and H$\alpha$. Both of
these ratios, however, depend on N/O, which can vary with O/H. A
significant fraction of the [\ion{N}{2}]~$\lambda 6584$ flux may also come
from the diffuse ionized ISM along the line of sight, and not from the
\ion{H}{2} region itself \citep{stasinska06}. Hence the N2 and O3N2 ratios
may be less suitable for providing a comparison between \Zemm\ and \Zabs .

Finally, we also include an estimate of \Zemm\ using SEL ratios calibrated
from photoionization models by \citet{kobulnicky04}. This index, which we
refer to as KK04 in Table~\ref{tab_HiiZ}, uses a combination of the well
known $R_{23}$ emission line ratio [([\ion{O}{2}]~$\lambda 3727$ +
[\ion{O}{3}]~$\lambda 4959, 5007$) / H$\beta$] and the ionization parameter
$q$ (the ratio of the flux of ionizing photons to the hydrogen
density). Their low-metallicity branch gives unique values of O/H providing
$\log$(O/H)$\la 8.5$, which, from the other SEL ratios given in
Table~\ref{tab_HiiZ}, appears to be true for \thisgal. While we have
avoided using the [\ion{O}{2}]~$\lambda 3727$ line for the reasons
mentioned above, \citet{lopez-sanchez12} have suggested that the KK04 index
is unique in being able to match values measured from the RL method. We
find the KK04 index for \thisgal\ to be between the Ar3O3 and S3O3 ratios,
and the ratios that use the [\ion{N}{2}] line, N2 and O3N2.

\section{COS OBSERVATIONS AND DATA REDUCTION}
\label{sect_cos}

Observations were made with COS using the $2\farcs 5$ diameter Primary
Science Aperture (PSA) and the G130M grating at Life Position 1. \qso\ was
observed for 6 orbits, broken into 2 visits, with total exposure times of
8192~s using the grating centered at 1291~\AA , and 8192~s when centered at
1327~\AA ; these two positions were chosen to provide some data in the gap
between the two segments of the photon-counting micro-channel plate
detector after coadding all the exposures \citep{green11}.
 
Data were processed with version 3.1.7 of the {\tt CALCOS} pipeline
software.  The post-processed coaddition of all the sub-exposures has been
discussed in detail in \citet{bowen16} and is not repeated here.  We
initially selected the QSO as a potential HST target given its GALEX FUV
flux listed in {\it General Release} GR4 as $104\pm 8 \mu$Jy. Subsequent
GALEX catalog releases, however, did not contain the QSO, only the
foreground galaxy. We measured a flux from the COS data of
$F_\lambda \text{(1400\AA )} = 3\times 10^{-16}$~\cgs, only a fifth of that
expected [equivalent to $F_\nu$(FUV)$\:= 20\:\mu$Jy in the GALEX FUV
band]. Either the QSO FUV flux is variable, or (more likely) the original
catalog over-estimated the flux due to additional light from the foreground
galaxy. The resulting S/N ratio of the spectrum was consequently lower than
expected, $\sim 4$ per (rebinned) 0.03~\AA\ pixel.

Comparison between features in the spectra obtained in the two HST visits
showed a clear shift of $\sim 5$ rebinned pixels, or $0.15$~\AA\ between
each. The spectrum from the second visit was shifted to match that of the
first, as the latter was found to show absorption lines from low-ionization gas
in the Milky Way that best matched the velocity of 21~cm emission features
seen in the Leiden/Argentina/Bonn (LAB) Survey of Galactic H~I
\citep{kalberla05}.
 
\begin{figure*}
\begin{center}
\vspace*{0cm}\hspace*{0cm}\includegraphics[width=18cm]{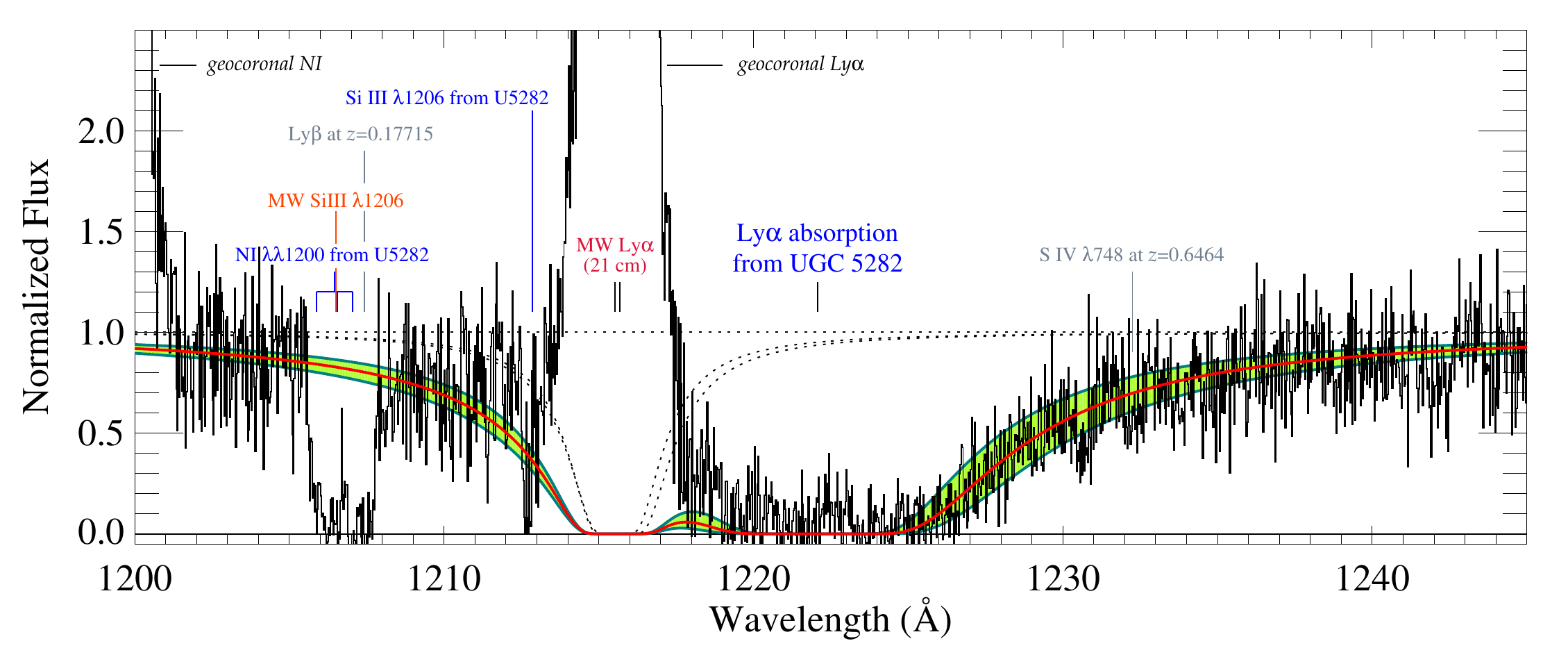}
\caption{Normalized COS spectrum of the QSO \qso .  
A composite \lya\ absorption line profile is shown as a red line, which is
comprised of absorption from UGC~5282  at
$v=1581$~\kms\ with $\log N$(H~I) = 20.89 ($+0.12, -0.21$),
and two components inferred from 21~cm emission line measurements (black dotted
lines) at $-35$ and 0~\kms\ with $\log N$(H~I) = 19.79 and 19.70,
respectively. 
The region shown in green corresponds to profile fits made to data
normalized by continuum fits that are 1$\sigma$ deviant from the
best fit continuum.
 The geocoronal emission lines from \lya\ and N~I are marked, as well as the wavelengths
of other detected absorption lines.
\label{fig_dla}}
\end{center}
\end{figure*}

\begin{figure}
\begin{center}
\vspace*{0cm}\hspace*{0cm}\includegraphics[width=0.45\textwidth]{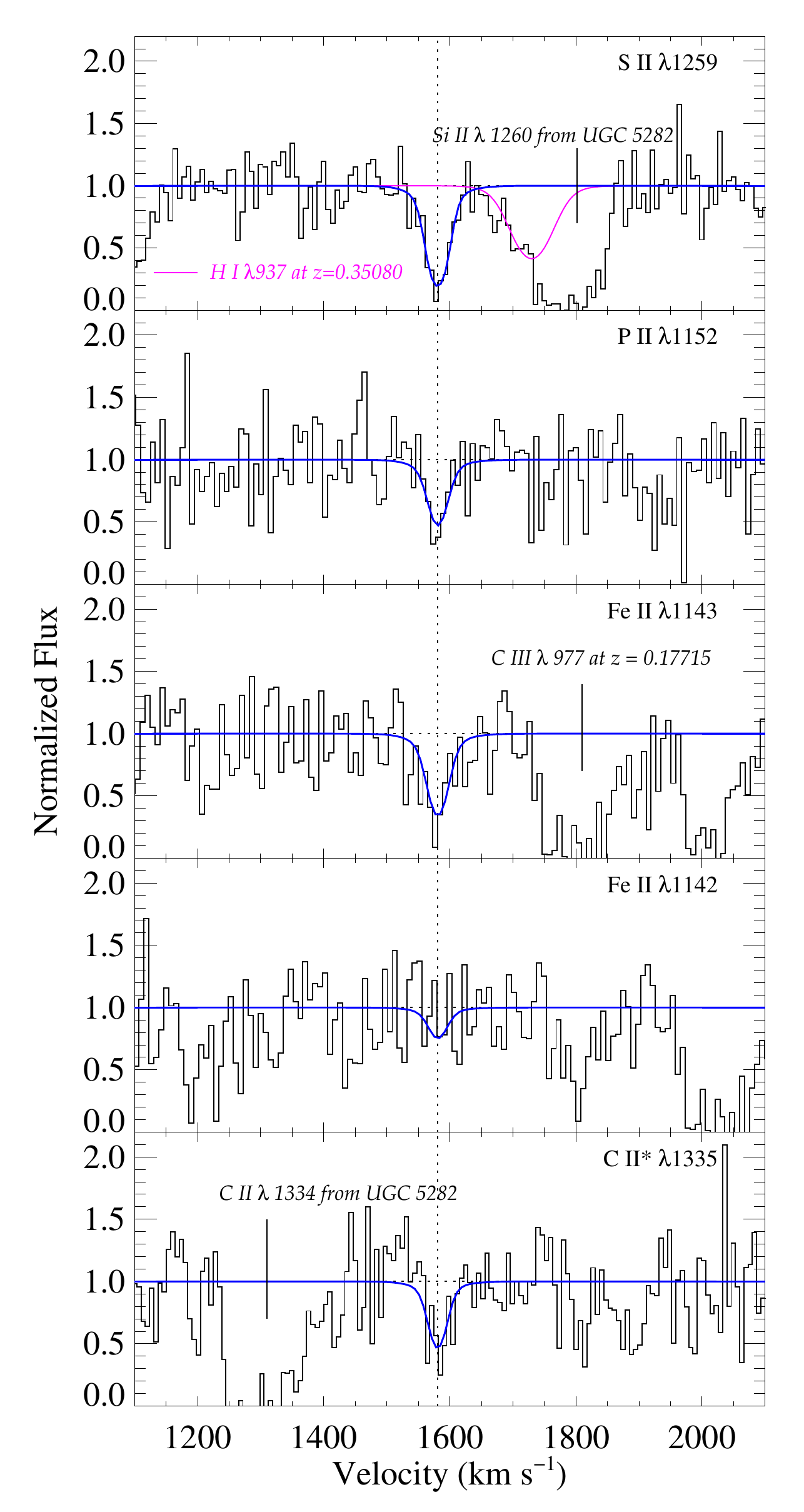}
\caption{Sections of the normalized COS spectrum of \qso\ featuring the set
  of relatively weak
  absorption lines from different ions at the velocity of
  \thisgal. 
Other significant intervening lines are also indicated.
\label{fig_weakstack}}
\end{center}
\end{figure}

Portions of the final co-added COS spectrum are shown in
Figures~\ref{fig_dla}$-$\ref{fig_check}, including the damped \lya\
absorption at the redshift of \thisgal\ (Fig.~\ref{fig_dla}), as well as
selected metal lines affiliated with the \ion{H}{1} absorption
(Figs.~\ref{fig_weakstack}$-$\ref{fig_check}).

\subsection{Line Identification and Profile Fitting}

The procedures for fitting the continuum of the spectrum, for measuring the
physical parameters of the detected absorption lines  --- their Doppler
parameters $b$, line of sight velocities $v$, and column densities $N$ --- 
and the methods used for determining errors in these values, are discussed
in detail in \citet{bowen08, bowen16} and are only summarized here. 

We normalized the final coadded spectrum by fitting Legendre polynomials
\citep[e.g.][]{sembach92} to areas free of features. Along with the best
fit continuum, we generated $\pm 1\sigma$ ``upper'' and ``lower'' error
``envelopes'' to represent the deviations that accompany the best fit. For
absorption lines of interest we constructed theoretical Voigt line profiles
from initial estimates of $v$, $N$, and $b$, and allowed these parameters
to vary until a minimum in $\chi^2$ between profile and data was reached.
Oscillator strengths for the lines were taken from \citet{morton03},
\citet{kisielius14} (\ion{S}{2}), \citet{federman07}
(\ion{P}{2}~$\lambda 1152$), or \citet{brown18}
(\ion{P}{2}~$\lambda 1301$).  Theoretical line profiles were convolved with
COS Line Spread Functions (LSFs) constructed by interpolating LSF
tables\footnote{Available online at
  \href{http://www.stsci.edu/hst/cos/performance/spectral_resolution/}{the
    Space Telescope Science Institute}.} to the relevant wavelength. Errors
to the parameters were calculated using a Monte Carlo approach, in which
400 synthetic spectra were constructed from the best fit profile using the
error arrays and refit to give new values of $v$, $N$ and $b$. These errors
were combined in quadrature with the differences between the best fit
values found when using the spectrum normalized by the upper and lower
continuum fits.

\subsection{Damped \lya\ Absorption in \thisgal \label{sect_dla}}

Figure~\ref{fig_dla} shows the \lya\ absorption from \thisgal. The line is
clearly damped, and blended with strong \lya\ from the Milky Way (MW).  In
order to define the velocity of the absorption system from \thisgal, we
first fitted 4 weak, low ionization lines that are expected to be
associated with high column density \ion{H}{1}: \ion{S}{2}~$\lambda 1259$,
\ion{P}{2}~$\lambda 1152$, \ion{Fe}{2}~$\lambda 1143$, and
\ion{C}{2}$^*\:\lambda 1335$, along with the part of the spectrum where
\ion{Fe}{2}~$\lambda 1142$ was expected, but not detected.  These metal
lines were fitted simultaneously, allowing $b$ and $v$ to vary, but
requiring that their final values be the same for each ion.  This produced
an absorption velocity of 1581~\kms\ that we used to fit the \lya\
absorption from \thisgal . These lines are discussed in more detail in
\S\ref{sect_weak} below.

Our model for the \lya\ absorption shown in Figure~\ref{fig_dla} consists
of $N$(\ion{H}{1}) from the MW as measured from two 21~cm emission
lines\footnote{Data taken from the
  \href{https://www.astro.uni-bonn.de/hisurvey/AllSky_profiles/index.php}{LAB
    survey} \citep{kalberla05}} at $-35$ and 0~\kms\ (shown as black dotted
lines in Fig.~\ref{fig_dla}), as well as the absorption from \thisgal . The
resulting blend of \lya\ lines (the red line in Fig.~\ref{fig_dla}, with
the uncertainty in the continuum fit shown in green) seems inadequate
between $\approx 1210-1212$~\AA , in that it appears to over-predict the
absorption in the blue wing of the DLA line; this may be because of a
contribution from flux in the wings of the geocoronal \lya\ emission line,
and/or a difference in the true $N$(\ion{H}{1}) along the line of sight
from that measured from the 21~cm emission lines, which are obtained from
observations that have a $0.6\degr$ beam-size. It is possible to obtain a
better fit in this wavelength range by reducing $N$(\ion{H}{1}) from the
MW, but the column density must be 2 dex smaller, with a value
$\sim 10^{18}$~\pcm\ compared to the $10^{20}$~\pcm\ measured at
21~cm. Such a value would be unusually low for the disk of the MW, and so
contamination from the geocoronal \lya\ emission line seems a more likely
explanation for the poor fit at these wavelengths.  However, beyond
$\sim 1221$~\AA, a damped \lya\ absorption line profile from \thisgal\ fits
the data well; evaluation of $N$(\ion{H}{1}) from the profile fitting is
dominated by data at these wavelengths, and is unaffected by whether
regions to the blue of the geocoronal \lya\ line are masked out or included
in the fit, or whether the velocity of the absorption from \thisgal\ is
allowed to vary or is fixed at the 1581~\kms\ discussed above. Our final
measurement of $N$(\ion{H}{1}) is given in Table~\ref{tab_weaklines}.

\begin{deluxetable*}{llccc}
\tablecolumns{5} 
\tablecaption{ Results from Voigt Profile Fits to Weak Absorption Lines
  from UGC 5282 \label{tab_weaklines} }
\tablehead{
\colhead{}
&\colhead{}
& \multicolumn2{c}{$v=1581 \pm 3$~\kms}
&\colhead{} \\
\cline{3-4}
\colhead{Ion}
&\colhead{Detected \tablenotemark{a}}
&\colhead{$b$}
&\colhead{}
&\colhead{}\\
\colhead{(X)}
&\colhead{Lines}
&\colhead{(\kms)}
&\colhead{$\log$[$N$(cm$^{-2}$)]}
&\colhead{$N$(X)/$N$(\ion{H}{1}) $-$ $Z_\odot$}
}
\startdata
\hline
\ion{H}{1}  & $\bm{\lambda 1216}$   &    ...   & 20.89 (+0.12,$-$0.21)    &  ... \\  
               &                                    &          &                                     &     \\
\multicolumn{5}{c}{Lines fitted simultaneously to give single $b$ and $v$ values}\\
\hline
\ion{S}{2}  & $\lambda 1250$,  $\bm{\lambda 1259}$ &   $15.5\pm5.9$  & 15.33 (+0.37, $-$0.15)  & $-0.82$ (+0.39, $-0.26$) \\
\ion{P}{2}  & $\bm{\lambda 1152}$                            &   $15.5\pm5.9$  & 13.65 (+0.27, $-$0.24)  & $-0.78$ (+0.30, $-$0.32) \\
\ion{Fe}{2} & $\bm{\lambda 1143}$, $\lambda 1144$, ($\bm{\lambda 1142}$)\tablenotemark{b} &   $15.5\pm5.9$ 
  & 15.02 (+0.33, $-$0.22) &  $-1.41$ (+0.35, $-$0.31)\\
\ion{C}{2}$^*$  & $\bm{\lambda 1335}$                     &   $15.5\pm5.9$  & 13.86 (+0.27, $-$0.25)  & ... \\
               &                                    &          &                                     &     \\
\multicolumn{5}{c}{Lines fitted assuming fixed $b$ and/or $v$}\\
\hline
\ion{O}{1} & $\bm{\lambda 1302}$                     &   15.5  &  16.51 (+0.49, $-$0.72)  & $-$1.14 (+0.51, $-$0.75) \\
               &                                                      & $29.8\pm17.7$\tablenotemark{c} & 15.15 (+0.55, $-$0.32) & .. \\
\ion{Si}{2} & $\bm{\lambda 1190, \lambda 1304}$ & 15.5   & 14.83$\pm0.36$  &   $-1.67$ (+0.38, $-$0.42) \\
               &                                                       & $30.0\pm 6.2$\tablenotemark{c} & 14.60 (+0.19, $-$0.12) & ... \\
\hline
\enddata
\tablenotetext{a}{Lines in bold indicate which lines were fit with Voigt profiles;}
\tablenotetext{b}{\ion{Fe}{2}~$\lambda 1142$ not detected, but data used to constrain fit;}
\tablenotetext{c}{Additional component at $v=1562$~\kms .}
\end{deluxetable*}

\subsection{Metal-line Column Densities from Weak Lines and Neutral Gas
  Metallicity \label{sect_weak}}

As discussed above, in order to derive column densities from the ISM of
\thisgal , we fitted theoretical Voigt profiles to a selection of weak,
unsaturated, metal lines. These are discussed below.

Of the three lines of the \ion{S}{2} triplet expected to be found from
\thisgal , \ion{S}{2}~$\lambda 1253$ is lost in the blend of
\ion{Si}{2}~$\lambda 1260$ and \ion{S}{2}~$\lambda 1259$ lines from the
Milky Way, while the \ion{S}{2}~$\lambda 1250$ line is contaminated by
\ion{H}{1}~$\lambda 930$ at $z=0.35080$ (we return to this line in
\S\ref{sect_check}).  \ion{S}{2}~$\lambda 1259$ is uncontaminated by other
absorption lines, and it provides the basis of the derivation of
$N$(\ion{S}{2}).

Absorption from \thisgal\ is also seen in a single line of
\ion{P}{2}~$\lambda 1152$, weak \ion{C}{2}*~$\lambda 1335$, and in several
\ion{Fe}{2} lines. Of the three lines of the
\ion{Fe}{2}~$\lambda\lambda 1143$ triplet, the $\lambda 1143.2$ line is
detected, and the $\lambda 1142.4$ line is absent. The strongest of the
three, $\lambda 1144.9$ is too strong given the strength of the other two
lines, and is likely contaminated by another feature that we cannot
identify. It is therefore not included in our fitting procedure.

The four lines (and one non-detection) used to derive the column densities
are shown in Figure~\ref{fig_weakstack}.  Fitting the lines resulted in a
Doppler parameter of $b=15.5\pm5.9$~\kms ; the profiles are shown in blue
in Figure~\ref{fig_weakstack}. All the parameters derived from the fit are
listed in Table~\ref{tab_weaklines}.

Sulphur is often taken as a proxy for oxygen when deriving gas-phase
metallicities, because the \ion{O}{1}~$\lambda 1302$ line which lies nearby
in the far-UV is saturated for high values of $N$(\ion{H}{1}), while weaker
\ion{O}{1} lines are at (shorter) wavelengths which are not always
available. Both O and S are $\alpha$-elements formed as part of the
$\alpha$-capture process in massive stars ($>10 M_\odot$) and released in
type~II SNae, and so are expected to track each other
closely. \citet{matteucci05} have pointed out that some S may also be
produced in type Ia SNae, but for Fe/H abundances of $ <\: -0.5$, the
difference in the O/S ratio from solar is $< 0.08$~dex, which, as we shall
see, is much less than the errors we derive in this paper. The other
problem in assuming that O and S abundances are the same is that the
relative O/S ratio may depend on the amount of depletion present. Oxygen is
only mildly depleted in the ISM, and varies between $0-0.3$~dex in both
short and long sightlines in the MW, and in the SMC \citep{kimura03,
  jenkins09, jenkins17}. \citet{jenkins09} introduced a line-of-sight
depletion factor $F_*$, which represents how far depletion has progressed
collectively for all elements along a sightline, such that a larger $F_*$
implies a stronger depletion for all elements. Although measuring S
abundances was less straightforward than for other species, Jenkins'
results indicate that the O/S ratio varies from $-0.15$ to $+0.4$~dex for
$F_*\:=\:0-1.0$. For very dense interstellar clouds, depletion may be far
more extreme \citep[e.g.][]{ruffle99}, but for diffuse clouds with only
small amounts of dust ($F_* \:\la \: 0.5$), using S/H to measure O/H is
likely a reasonable assumption.

In addition, \ion{S}{2} is not only expected to be the dominant sulphur ion
in neutral gas with $N$(\ion{H}{1}) as high as that seen towards \thisgal,
but any ionization correction necessary to calculate $N$(S) is expected to
be negligible. This is not only true for gas ionized by the extragalactic
UV flux \citep[e.g.][]{sbs1543_2} but also for ionization by \ion{H}{2}
regions \citep[e.g.][]{james14}. The same arguments also apply to
\ion{P}{2}, which we discuss below.

If we therefore assume that all the sulphur is in the form of \ion{S}{2},
and that S is undepleted, then (S/H)$\:=\:$(\ion{S}{2}/\ion{H}{1}), and the
metallicity $Z$ of the neutral gas in \thisgal\ is

\begin{equation*}
Z_{\text{H I}}(\text{S}) = \left[ \frac{\text{S}}{\text{H}} \right] \equiv 
\log \left( \frac{\text{S}}{\text{H}} \right) -  \log \left(
  \frac{\text{S}}{\text{H}} \right)_\odot =  -0.82\: (+0.39, -0.26)
\end{equation*}

\noindent where $12+\log\, (\text{S/H})_\odot\:=7.26\pm 0.04$ is the
protosolar  value from \citet{lodders03}.

P is another element that is assumed to track O and S. Although P is not an
$\alpha$-element, it is thought to be produced in the same massive stars
that produce O and S \citep{cescutti12}, and so follows the
$\alpha$-elements as they are dispersed by type~II SNae. In the Milky Way,
P and S deplete by very similar amounts as $F_*$ varies
\citep{jenkins09}. We measure

\begin{equation*}
Z_{\text{H I}}(\text{P}) = \left[ \frac{\text{P}}{\text{H}} \right] =
-0.78\: (+0.30, -0.32),
\end{equation*}

\noindent again, assuming no depletion and no ionization corrections, and using
$12+\log\, (\text{P/H})_\odot\:=5.54\pm 0.04$ as the solar value. 

The metallicities of S and P derived are sufficiently similar within their
errors that we can take a simple weighted average of the two values to give a
final estimate of the metallicity:

\begin{equation*}
Z_{\text{H I}}(\text{S,P}) \simeq -0.80 \pm 0.24\:.
\end{equation*}

\noindent i.e. $0.16\pm0.09$ times, or $\approx 1/6$ of, the solar value.

Finally, our measurements of Fe give an indication of the depletion $\Delta$
in the gas, since iron is known to be strongly depleted in diffuse
clouds. If (\ion{Fe}{2}/\ion{H}{1}) is a good measure of Fe/H, then 

\[
\Delta\text{(Fe)} = \text{[Fe/H]} - Z_{\text{H I}}\text{(S,O)} = -0.61 \pm 0.42 .
\]

\noindent This value is much smaller than the depletion of Fe relative to S in the MW
and the LMC --- for all values of $F_*$ --- but is entirely consistent with the depletions
found in DLAs at higher redshift \citep[e.g.][]{kulkarni15}.

\subsection{Consistency checks \label{sect_check}}

\begin{figure}
\begin{center}
\vspace*{0cm}\hspace*{0cm}\includegraphics[width=8.5cm]{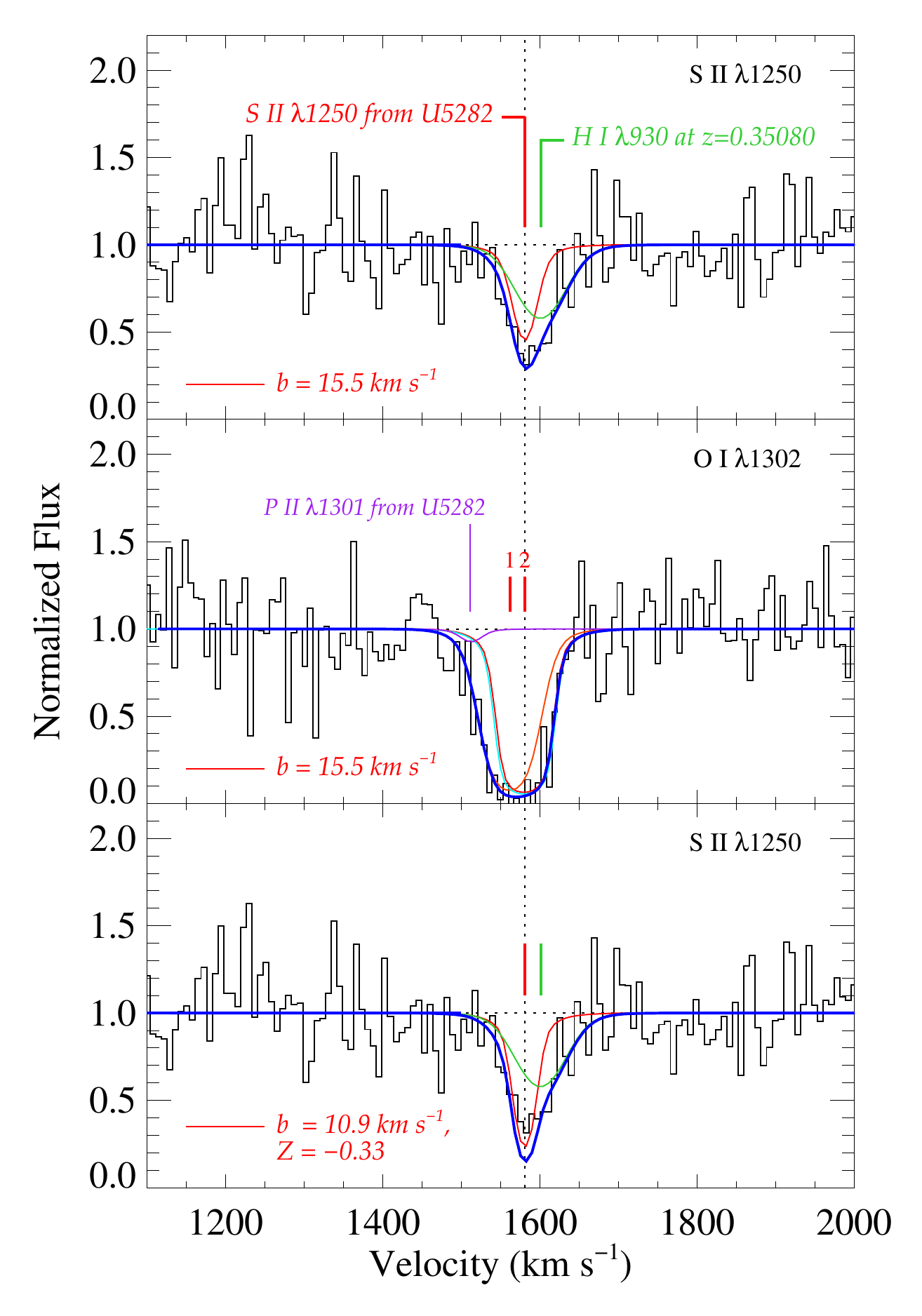}
\caption{Predicted profiles of \ion{S}{2}~$\lambda 1250$ and \ion{O}{1}~$\lambda 1302$
  lines arising from \thisgal. {\it Top panel}: the theoretical line profile for the \ion{S}{2}~$\lambda
  1250$ line assuming $N$(\ion{S}{2}), $v$ and $b$ derived from the \ion{S}{2}~$\lambda
  1259$ line, given in Table~\ref{tab_weaklines}, is shown as a red
  line. The \ion{S}{2} line is blended with
  a higher redshift \ion{H}{1} Lyman series line (predicted as the green line
  given the measured \ion{H}{1} and $b$-value from other Lyman lines at the
  same redshift), which makes the final \ion{S}{2} line profile
  uncertain, but the predicted blend of the \ion{S}{2} and \ion{H}{1} lines,
  shown in blue, matches the data well. {\it Middle
  panel}: a theoretical line profile fit for the \ion{O}{1} line
  requires two components, assuming that one of the components (labelled '2' in this
  figure) has 
  the same $v$ and $b$ values given in Table~\ref{tab_weaklines}. The two
  components are shown as red lines, and their composite blend in blue. Although
  the fit appears to match the data well, component 2 is saturated and
  insensitive to  $N$(\ion{O}{1}): the cyan line shows the
  predicted line profile of component 2 when it has the value of
  $N$(\ion{O}{1}) expected given the same metallicity $Z_{\text{H
      I}}$  derived from the \ion{S}{2}~$\lambda
  1259$ line, which is 0.3 dex smaller than the best-fit line profile shown
  in red. The two profiles are largely indistinguishable. 
Also shown in purple is the predicted
  profile of the \ion{P}{2}~$\lambda 1301$ line based on the values of the
  \ion{P}{2}~$\lambda 1152$ line given in Table~\ref{tab_weaklines}.
{\it Bottom panel:} the same \ion{S}{2}~$\lambda
  1250$ line shown in the top panel, but showing a profile for an absorbing
  cloud that has the same metallicity as the central \ion{H}{2} region,
  \Zemm (O)$\: =\: -0.37$, which requires $b=10.9$~\kms . The profile is
  clearly a poorer match to the data than the best-fit model shown in the
  upper panel, although the difference is small.  
  \label{fig_check}}
\end{center}
\end{figure}

The spectrum of \qso\ contains several other absorption lines from
\thisgal\ that are less suitable for abundance determinations, as they are
either saturated or blended with unrelated features. However, we can check
whether the model constructed for the weak lines shown in
Figure~\ref{fig_weakstack} and listed in Table~\ref{tab_weaklines} can
successfully predict the profiles of these other lines.

For example, the top panel of Figure~\ref{fig_check} shows the
\ion{S}{2}~$\lambda 1250$ absorption arising from \thisgal. This line was
not used to constrain any column densities or $b$-values as the line is
contaminated by \ion{H}{1}~$\lambda 930$ at $z=0.35080$. In this system,
Ly$\beta$, Ly$\gamma$ and Ly$\delta$ are all detected in the COS spectrum,
enabling us to calculate the expected profile for the $\lambda 930$ line at
the position of the \ion{S}{2} line.  We find
$\log N$(\ion{H}{1})$\:=\:15.55 \pm 0.07$ and $b=37.4\pm 3.9$~\kms\ for
this high-$z$ system. The profile is shown as a green line in the figure.
The \ion{S}{2}~$\lambda 1250$ line itself can be predicted using the values
of $b$, $v$ and $N$(\ion{S}{2}) derived from the \ion{S}{2}~$\lambda 1259$
line (Table~\ref{tab_weaklines}), and this is shown as a red line in
Figure~\ref{fig_check}. The blend of the two is shown in blue, and the
results appears to agree well with the data.

The middle panel of Figure~\ref{fig_check} shows the
\ion{O}{1}~$\lambda 1302$ absorption from \thisgal . \ion{O}{1} is an
important species to observe, because O shows much less depletion as a
function of $F_*$ than S or P, and the ionization corrections needed to
convert from \ion{O}{1} to O are negligible, since charge exchange locks
the ionization of oxygen to that of hydrogen \citep{field71}.  The
\ion{O}{1}~$\lambda 1302$ absorption from \thisgal\ is saturated and very
insensitive to the value of $N$(\ion{O}{1}), but if we assume that there
must exist a component with the same $b$ and $v$ values as the weak lines
listed in Table~\ref{tab_weaklines}, then we find that the line must be
comprised of at least two components, and that with $b$ and $v$ fixed for
the DLA component (labelled `2' in the figure), a best fit can be computed.
For completeness, the resulting values are listed in
Table~\ref{tab_weaklines}; the errors on $N$(\ion{O}{1}), and hence on the
metallicity, are large, but the value of $Z$(O) is at least consistent with
the values of $Z_{\text{H I}}(\text{S})$ and $Z_{\text{H
    I}}(\text{P})$. Figure~\ref{fig_check} also shows the predicted line
profile of component 2 for the same fixed values of $b$ and $v$, but with
$N$(\ion{O}{1}) derived from the same metallicity as inferred for sulphur,
$Z_{\text{H I}}(\text{S}) = -0.80$. This is shown as a cyan line in the
figure; it is barely distinguishable from the best fit for component 2,
which is 0.3 dex larger in column density. This demonstrates how
insensitive the \ion{O}{1}~$\lambda 1302$ is for measuring $N$(\ion{O}{1}),
but also that a simple 2 component model with a priori knowledge of some of
the fitting parameters can at least reproduce the data.

%
%
%
%
%
%
%
%
%
%

\subsection{\lya\ Emission from \thisgal \label{sect_lyaemm}}

\begin{figure}
\begin{center}
\hspace*{-0.5cm}\includegraphics[width=0.5\textwidth]{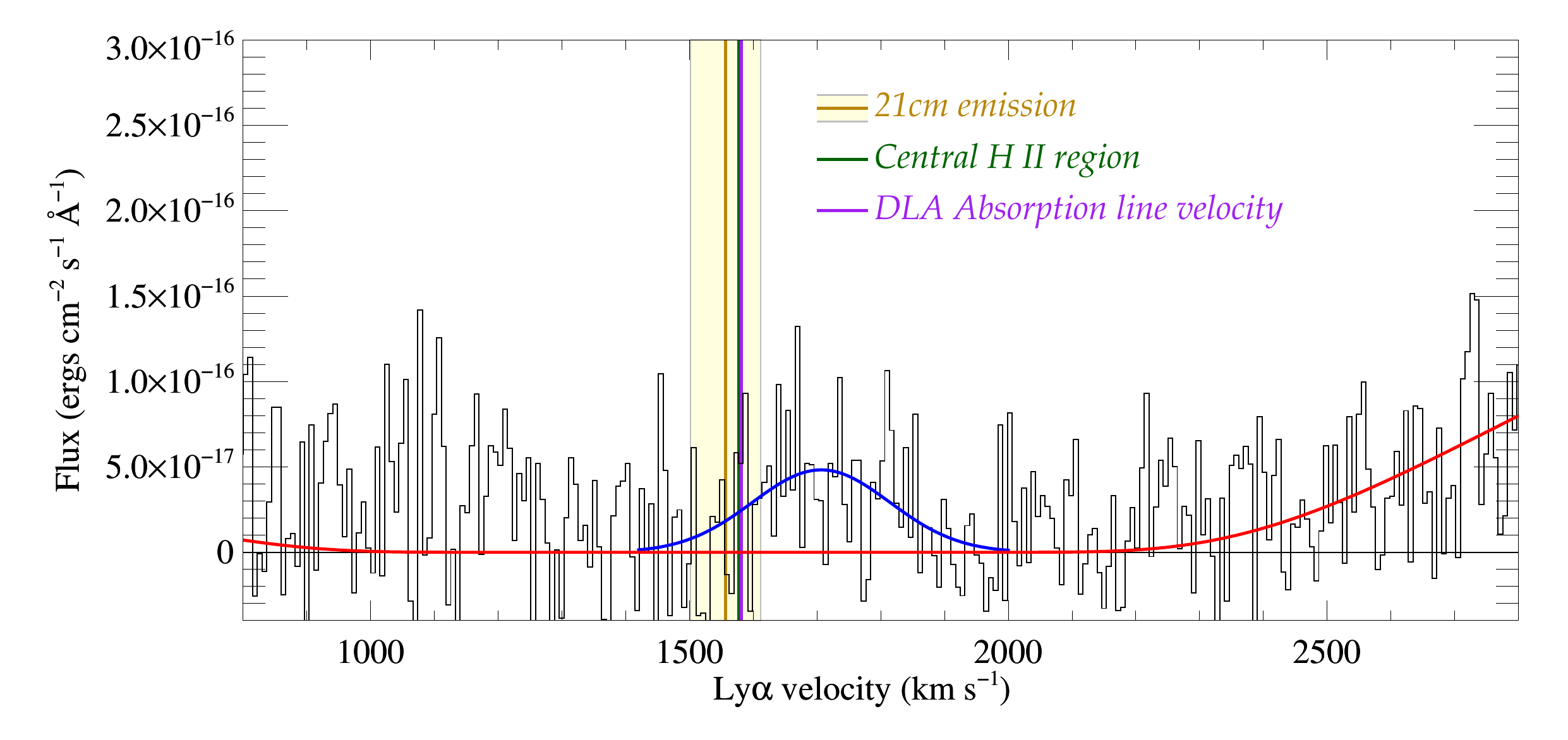}
\caption{Comparison of \lya\ emission towards \qso\ with three estimates of
  the velocity of \thisgal . The yellow region shows the 21~cm emission
  tabulated by \citet{schneider90} with a width equivalent to $W_{50} =
  110$~\kms , the
  width of the \ion{H}{1} profile at 50\% of its peak. The redshift of the
  central \ion{H}{2} measured by SDSS is shown in green, and the velocity
  of the DLA component at 1581~\kms\ is shown in purple. The red line
  indicates the \lya\ absorption profile. Although we consider the emission
  at 1706~\kms\ to be real, an apparent excess of flux between
  900$-$1200~\kms\ may be due to contamination from geocoronal \lya\
  emission.
\label{fig_lyaemm}}
\end{center}
\end{figure}

\lya\ is detected in emission at the bottom of the damped \lya\ absorption
line (Fig.~\ref{fig_lyaemm}). The emission is clearly redward of the
absorption lines seen towards \qso, as well as the systemic velocity of
\thisgal\ measured from 21~cm emission lines. The flux integrated over a
Gaussian line fit is $5.2 \pm0.3 \times 10^{-17}$~\cgs, centered at
1706~\kms\ with a width of $\sigma = 108$~\kms .  The corresponding
luminosity is $3.0 \pm 0.2 \times10^{36}$~ergs~s$^{-1}$, although this only
represents the amount seen through the COS aperture. There may also be
emission blueward of the galaxy's systemic velocity ($\sim 900-1200$~\kms
); however, this may be directly related to the flux from the red wing of
the wide geocoronal \lya\ emission (see Fig.~\ref{fig_dla}), which makes it
difficult to claim that the emission is real. The positive offset in
velocity of the emission from \thisgal\ might suggest that either the
emitting gas lies between us and the galaxy and is falling into it, or else
the gas is outflowing from the other side of galaxy. Unfortunately, neither
scenario is supported by the detection of any high velocity absorption
components.

Alternatively, the emission could be a radiative transfer effect due to
resonant scattering of the \lya\ line from the galaxy's ISM. The offset in
the \lya\ velocity is, for example, reminiscent of that seen in nearby
galaxies \citep[e.g.][and refs.~therein]{orlitova18, hayes15}. In this
case, the \lya\ emission characteristics could have implications regarding
the escape of Lyman continuum emission from the galaxy
\citep[e.g.][]{dijkstra16} and feedback effects
\citep[e.g.][]{mcKinney19}. The low S/N of our data and the likely presence
of geocoronal \lya\ contamination at lower velocities preclude a detailed
analysis of these issues, but higher S/N follow-up observations and mapping
of the global \lya\ emission from \thisgal\ could provide valuable insights
on these topics.

\bigskip
\section{Differences in \Zabs\ and \Zemm\ in \thisgal\ and other
  star-forming galaxies}
\label{sect_delta}

With both \Zemm\ and \Zabs\ measured for \thisgal , the difference between
the two is simply

\[
\delta\text{(X)} = Z\text{(X)}_{\text{H~II}} - Z\text{(X)}_{\text{H~I}}\:.
\]

We take \Zemm (O)$\: =\: -0.37\pm0.10$ from
Table~\ref{tab_HiiZ} (the S23 strong line calibration, and assuming an
error of 0.10 in the calibration) and \Zabs (P,S)$\:=-0.80\pm
0.24$; this gives 

\[
\delta(\text{P,\,S}) = 0.43\pm0.26\: , 
\]

\noindent a value that indicates that the \ion{H}{2} region abundances are
a factor of $\sim 3$ higher than those in the outer diffuse ISM.

How robust are the column densities listed in Table~\ref{tab_weaklines} and
the resulting value of $Z\text{(X)}_{\text{H~I}}$? Uncertainties arising
from the modest S/N of our data are largely reflected in the large
uncertainties attached to the measurements. In addition, the leverage that
the available lines provide to measure column densities independent of
their $b$ values is not strong: if lines are unresolved, column densities
are only robust for sets of lines with different transition probabilities,
and for the spectrum of \qso, we only have two lines of the
\ion{S}{2}~$\lambda 1256$ triplet available, one of which is contaminated
by an intervening \ion{H}{1} line at a higher redshift. This makes the
model constructed for the absorption more uncertain.

One simple test of whether $\delta$(P,S) is really different from zero is
to re-fit the absorption lines assuming that they have the {\it same}
metallicity as the gas in the \ion{H}{2} region. With $v$ and $N$ fixed for
the \ion{S}{2} and \ion{P}{2} lines, we find that the same lines shown in
Figure~\ref{fig_weakstack} can be best fit with a single component that has
$b=10.9$~\kms. However, not only is $\chi^2$ worse for the fits with these
assumed column densities, but both the \ion{S}{2}~$\lambda 1259$ and
\ion{P}{2}~$\lambda 1152$ lines seem too strong compared to the data. The
bottom panel of Figure~\ref{fig_check} shows the difference for the
\ion{S}{2}~$\lambda 1250$ line.  The figure demonstrates  that although the
difference is small, the better fit to the data is still that seen in the
top panel which gives the values of $N$(\ion{S}{2}) and $b$ listed in
Table~\ref{tab_weaklines} and the lower value of
$Z\text{(S,P)}_{\text{H~I}}$. While we caution that a definitive model for
the absorption is difficult to define with the current data, we assume that
the value of $\delta$(P,S) given above is a reasonable estimate.

In order to compare the values of $\delta$(P,S) from \thisgal\ with
sightlines to other extragalactic star forming regions, we compiled a list
of low-$z$ galaxies where both emission and absorption lines have been
measured towards individual star clusters. The galaxies selected are shown
in Table~\ref{tab_collection}. We only included sightlines where
$\log N$(\ion{H}{1})$>20.1$ in the neutral gas absorption, in order to
remove uncertainties from unknown ionization corrections in converting
$N$(S) and $N$(P) from $N$(\ion{S}{2}) and $N$(\ion{P}{2}).

A significant problem in compiling these data, however, is knowing which
published value of O/H should be used in calculating \Zemm\ for any
particular galaxy.  There are
multiple ways to measure O/H from emission lines; in particular, abundances
determined from RL methods are known to be $0.2-0.3$ dex higher than those
measured from the $T_e$ method, while the former method agrees better with
the metallicities of individual stars associated with \ion{H}{2} regions
\citep[e.g.][and refs.~therein]{lopez-sanchez12}. Many of the published
values of O/H for our selected galaxies were constructed using the $T_e$
method, values of which may be inappropriate for \thisgal\ whose
metallicity was measured using SEL methods (Table~\ref{tab_HiiZ}). To
provide a fair comparison therefore, we calculated new SEL values of $\log$
O/H for our set of galaxies using the same indices we used for \thisgal,
namely Ar3O3, N2, O3N2 and KK04. To do this, we used the tables of emission
line intensities provided in the published papers listed at the end of
Table~\ref{tab_collection}. In most cases, we used line intensities that
were already corrected for dust absorption and underlying Balmer absorption
by the authors; although these corrections may have been made using, e.g.,
different extinction laws or $R_V$ values, the differences are small in the
optical for star-forming galaxies, and as we discuss below, are
insignificant compared to the systematic calibration errors in the
metallicities derived from emission line ratios. Four of the selected
galaxies also had SDSS spectra available, which enabled us to measure
\Zemm\ in exactly the same way as we did for \thisgal, following the same
analysis described in \S\ref{sect_h2}.

For the galaxies listed in Table~\ref{tab_collection}, many authors report
line intensities for multiple positions within a galaxy. We discuss in
\S\ref{sect_discussion} what is currently known about the variations in
\Zemm\ with position for these types of star-forming dwarf galaxies, but in
order to avoid any errors that might arise from spatial variations in
emission line metallicity, we only included data if they were obtained at
the same position as the apertures used for the absorption line
measurements.  In most cases, the regions selected for both absorption and
emission line measurements were simply the brightest region of a galaxy,
usually at its center. Similarly, for the new measurements made from
existing SDSS spectra, in three of the four cases the spectroscopic fibers
appear to have been positioned on the same regions used to record the
emission lines. For emission line measurements published several decades
ago, no information is given on the exact positions of the slits used, and
we assume that the observations were made of the brightest regions of a
galaxy. In these cases, we list in column 4 of Table~\ref{tab_collection}
an indication of the spatial scale of the spectroscopic observations, if
given by the authors.  We list the emission line ratio metallicities
measured at any and all positions within the absorption line aperture in
columns 5$-$8 of Table~\ref{tab_collection}, in order to show the possible
variations in $\log$(O/H) that exist along the same sightlines where
absorption line metallicities are measured.

For the absorption line measurements, \Zabs(O), \Zabs(S) and \Zabs(P) were
calculated from the published column densities and using the same solar
abundance used for \thisgal . These are listed in columns 11$-$13 of
Table~\ref{tab_collection}. In calculating $\delta$, we favored \Zabs(S)
over \Zabs(P), and used \Zabs(O) only once (NGC~1705) where no \ion{S}{2}
or \ion{P}{2} lines had been measured. The large differences in \Zabs(O)
compared to \Zabs(S) and \Zabs(P) seen towards several objects almost
certainly arise from the fact that  a single, likely saturated, O~I
absorption line was used. 

\begin{figure}
\begin{center}
\hspace*{-0.5cm}\includegraphics[width=0.5\textwidth]{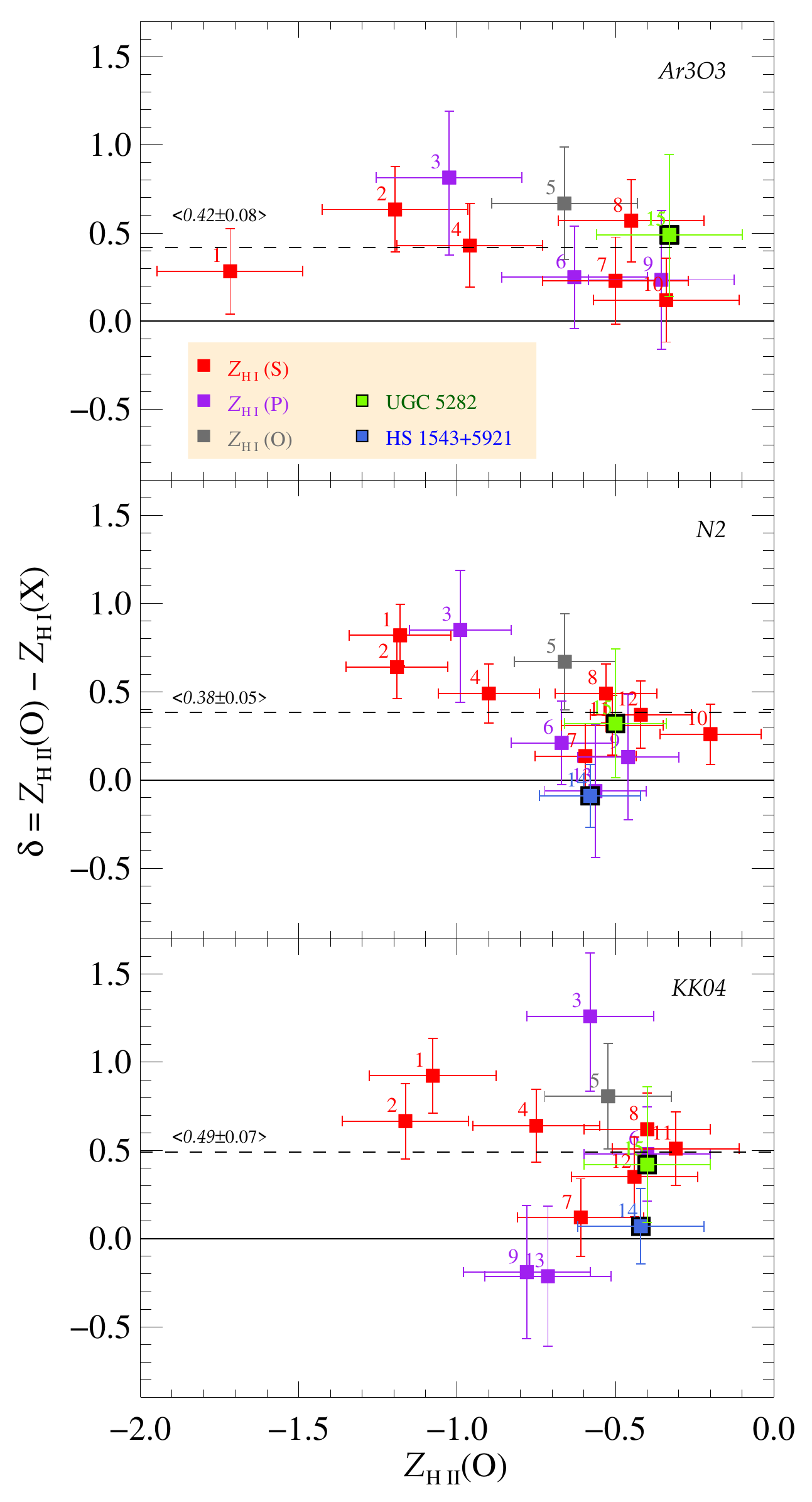}
\caption{Plots showing the difference in metallicities $\delta$ between ionized gas
($Z_{\text{H~II}}$) 
measured from emission lines from \ion{H}{2} regions, and neutral gas
($Z_{\text{H~I}}$) measured by absorption lines towards the \ion{H}{2}
regions, or, in two cases, background QSOs. 
To first order, \Zemm\ represents the metallicity of gas within or close to
the \ion{H}{2} region, while \Zabs\ measures the abundances of the ISM in
the rest of the galaxy along the line of sight.
\Zemm\ is measured using 3
different methods, the Ar3O3, N2, and KK04 strong emission line
ratios. Points are numbered with the
IDs listed in column 1 of Table~\ref{tab_collection}. The use of \ion{S}{2}, \ion{P}{2}
or, in a single case, \ion{O}{1} absorption lines for measuring \Zabs\  are indicated by the use of red, purple,
or gray squares, as described in the legend in the top panel; when
measuring $\delta$, \Zabs (S) was preferred over \Zabs(P), both of which
were selected over \Zabs(O). Values of $\delta$ for the 2 QSOs intercepting
foreground galaxies are also indicated. Dashed horizontal lines indicate
weighted averages of all the values of $\delta$. For each panel, the error bars are a quadratic sum of the
errors in the column densities from measuring \Zabs , and the systematic
calibration errors in determining \Zemm\ given by $\sigma_{\text{SL}}$ in Table~\ref{tab_HiiZ}.
\label{fig_delta}}
\end{center}
\end{figure}

Values of $\delta$ are shown in Figure~\ref{fig_delta}, where $\delta$ for
an individual object is the mean of the values listed for that object in
Table~\ref{tab_collection}.  The figure excludes the values of $\delta$
from the O3N2 method, since the index is known to break down for
metallicities below 12+log(O/H)$\: \la\: 8.3$
\citep{pettini04,marino13_ra}, and all the galaxies in
Table~\ref{tab_collection} have O3N2 less than this limit. For the other
indices, the dominant contribution to the errors in $\delta$ is that from
the absolute calibration of Ar3O3, N2, and KK04, and we take a conservative
approach and simply use the same values of $\sigma_{\text{SL}}$ listed in
column 3 of Table~\ref{tab_HiiZ} as the error in \Zemm\ for each
object. That is, we plot $\delta = \:(<$\Zemm$>\: -
\:\:$\Zabs)$\:\pm \sigma_{\text{SL}}$ for each galaxy in the figure.  These
errors are significantly larger than the errors in individual emission line
measurements, as well as the resulting errors in the emission line ratios,
and are larger than most of the errors in \Zabs(X), although the latter are
still added in quadrature to the former.

Also plotted in Figure~\ref{fig_delta} are the values of $\delta$ for the
two dwarf galaxies that are probed by QSO sightlines, SBS~1543+5921, and
our new results for \thisgal . The value of $\delta$ for \thisgal\ is
higher than that seen towards SBS~1543+5921, although their errors clearly
overlap. Both values of $\delta$ are similar to the ensemble of $\delta$
values for the star forming galaxies. Weighted averages for all the
sightlines are shown as dashed lines in the figure, and are of a similar
value, $<\delta> = 0.4-0.5$. This is a factor of 2 higher than the value
of $<\delta> = 0.20\pm0.23$, found by \citet{james18} using a smaller
number of sightlines. The difference between our value and theirs is within
the range of systematic offsets found when different indices are used to
measure emission line metallicities.

\section{DISCUSSION}
\label{sect_discussion}

The values of $\delta$ shown in Figure~\ref{fig_delta} suggest that a real
difference exists between the metallicity of the ionized ISM in which stars
are forming (\Zemm) and the metallicity of the neutral ISM that constitutes
the galaxy (\Zabs). The suggestion that $\delta$ is greater than zero
towards \ion{H}{2} sources in nearby starburst galaxies is not new
\citep[e.g.][]{cannon05,lebout13,james18}, but the sightline to \qso\
measures the bulk of \thisgal 's ISM metallicity $\sim 1$~kpc {\it away}
from the brightest regions of star formation and along a sightline through
the {\it entire} galaxy.  If a non-zero value of $\delta$ is a common
feature in low mass galaxies, then our result may indicate that there
exists a `baseline' galaxy metallicity that is better measured from its
cool neutral ISM. The true `metallicity' of a galaxy is not that measured
from \Zemm\ but from \Zabs .

Why should \Zabs\ and \Zemm\ be different?  Star forming regions certainly
add metals to the ISM, but the metallicity measured is that of gas enriched
from earlier episodes of star formation, and not from the bursts that are
marked by the \ion{H}{2} regions we observe. The \ion{H}{2} regions that
are currently active are ejecting material through supernovae or stellar
winds, and much of the gas that contains the newly made metals is expected
to be hot, at temperatures of $10^{6-8}$~K, at least for a few hundred Myr
\citep{emerick18}. The temperature of the gas from which \Zemm\ is measured
now is less than this, $10^{4-5}$~K \citep[e.g.][]{kewley19}, and new
metal-rich gas contributes little to the optical line emission used to
measure \Zemm . Evidence for the hot gas can be found in its X-ray emission
\citep[][and refs.~therein]{mcquinn18} and in \ion{O}{6} absorption in the
FUV \citep{grimes09}.  Thus the `enriched' material whose metallicity is
measured by \Zemm\ is gas that was once hot but has cooled and been able to
return to the ISM, and is now ionized by nearby stars.  This re-enrichment
of the ISM may not be particularly efficient; the outflows remain hot for
at least the lifetime of the \ion{H}{2} regions, several tens of Myr
\citep{legrand01}, and may be sustained over the lifetime of the starburst
activity, perhaps as much as several hundred Myr
\citep{mcquinn18}. Complete recycling of gas may take up to a Gyr
\citep{christensen16}. In addition, metals may be lost to the IGM because
of the low gravitational potential of the dwarf galaxy
\citep[e.g.][]{MacLow99, ferrara00, mcquinn15, emerick19} although how
much material (in either mass or metals) escapes remains unclear
\citep{muratov17,McQuinn19}.

Although the exact details of how gas cools and remixes with the ISM are
complex, the enrichment is likely to be spread over the size of the
outflows, i.e., over kpc scales. Indeed, there is good evidence for a well
mixed ISM in low mass galaxies. The most recent observations using multiple
spectroscopic slits or integral field spectrographs suggest that on kpc
scales, gas metallicity is quite homogeneous, with either no obvious
variations in \Zemm\ between sources
\citep[e.g.][]{cairos17,lagos16,kehrig16,lagos12}, or only small
perturbations characterized by weak metallicity gradients \citep[e.g.][and
refs.~therein]{bresolin19,annibali19, annibali17,
  annibali15}. Consequently, even if \thisgal\ had a metallicity gradient
similar to other low mass galaxies, the difference in \Zemm\ over the 1~kpc
distance between the center of \thisgal\ and the sightline to \qso\ ought
to be negligible if gas is well mixed.

Moreover, if mixing is effective over kpc scales, then we would expect the
neutral gas observed towards the QSO and the ionized gas seen towards the
central \ion{H}{2} region to have the same metallicity. In which case, in
order to find \Zabs$\:<\:$\Zemm, a significant fraction of the neutral gas
must be many kpc away from the star forming regions in order to not be
contaminated.  Such an explanation was adopted by \citet{cannon05}, who
postulated the existence of a low metallicity halo beyond the inner ISM to
explain the discrepancy between \Zabs\ and \Zemm\ towards star forming
regions in NGC~625. Our results are consistent with this idea, and provide
additional evidence for IGM gas feeding into galaxies through streams from
the cosmic web \citep[e.g.][and refs.~therein]{sanchez-almeida14a}.  Dwarf
galaxy {\it pairs} in particular are thought to have enhanced star
formation because they are fed by significant reservoirs of neutral gas in
which they reside, or because of their mutual interactions
\citep{lelli14,stierwalt15, pearson16}. As noted in \S\ref{sect_environs},
\thisgal\ does lie near a galaxy of similar mass (UGC~5287 in
Fig.~\ref{fig_lss}) and interactions between the two galaxies may be the
source of gas flowing into \thisgal . For example, \citet{pearson18} have
suggested that multiple encounters between dwarf galaxies can `park' gas at
significant distances from the protagonists, which can then return over
several Gyr. It remains unknown whether the metallicity of such returning
debris would be low enough to cause the decrease in metallicity in either
\thisgal\ or UGC~5287, but it may reflect whatever build-up of metals
occurred in the dwarfs at much earlier times. Alternative tidal models that
cause strong metallicity gradients and the {\it removal} of low metallicity
gas at the edges of dwarf galaxies \citep[][]{williamson16} seem less well
supported by our results.  Indeed, the situation for \thisgal\ may be even
more complicated, depending on whether it has entered the halo of NGC~3003
(Fig.~\ref{fig_lss}) and has begun to feel any effects from ram-pressure
stripping by gas in the host's halo. The effects on the metallicity of the
dwarf galaxy may, however, be less significant than any effects from tidal
stripping \citep{williamson18}.

Although dwarf galaxies usually show a well mixed ISM, there are
exceptions. In particular, some dwarf galaxies demonstrate regions where
\Zemm\ decreases significantly ($\sim 1$ dex over only a few hundred pc) at
positions of high star formation surface density compared to \Zemm\ from
the rest of the galaxy, \Zhost\ \citep{richards14,sanchez-almeida18b,
  sanchez-almeida15,sanchez-almeida14b}. This is again attributed to the
influx of low metallicity gas from the IGM that dilutes the ISM, but this
time on sub-kpc scales.  These examples argue for much more local mixing of
IGM gas with the ISM.  \Zabs\ towards these specific regions of low \Zemm\
has not been measured, and whether \Zabs\ would be even lower than \Zemm\
is not known, and likely depends on the degree of mixing.  In principle, a
similar scenario might exist in \thisgal, with the central \ion{H}{2}
region being the site of inflowing gas. This is hard to reconcile with our
result that \Zabs$\:<\:$\Zemm\ though; the inflow towards the \ion{H}{2}
region might occur on sub-kpc scales, yet we see a lower metallicity along
the QSO sightline a kpc away, where there is --- by definition in this
model --- less inflow.  The obvious test of such local mixing for \thisgal\
would require a measurement of \Zabs\ towards the central \ion{H}{2}
region, and a measurement of \Zemm\ towards the QSO line of sight: for only
local mixing, \Zabs\ towards the \ion{H}{2} region would be less than
\Zabs\ towards the QSO as low metallicity gas mixes with the ISM, while
\Zemm\ towards the QSO would be higher than \Zemm\ towards the central
\ion{H}{2} region.

The difference in the baseline metallicity that is assumed for a galaxy can
be important for understanding the evolution of galaxies. For example, in
the canonical closed-box model of a galaxy, the metallicity can be
expressed as a function of the gas mass fraction
[$M_{\text{gas}}/(M_{\text{gas}}+M_{\text{stars}})$] and the yield [the
ratio of the rate at which metals are ejected into the ISM to the rate at
which hydrogen is removed by star formation \citep{searle72}]. Deviations
of the measured metallicity of the gas compared to expected values are
usually taken to indicate the presence of either inflowing or outflowing
gas \citep[e.g.][and refs.~therein]{lagos18}. A difference of
$\sim 0.5$~dex in the assumed `true' metallicity of the gas can change
entirely whether a measured metallicity is consistent with flows of gas in
or out of a galaxy.

\begin{figure}
\begin{center}
\hspace*{-0.5cm}\includegraphics[width=0.5\textwidth]{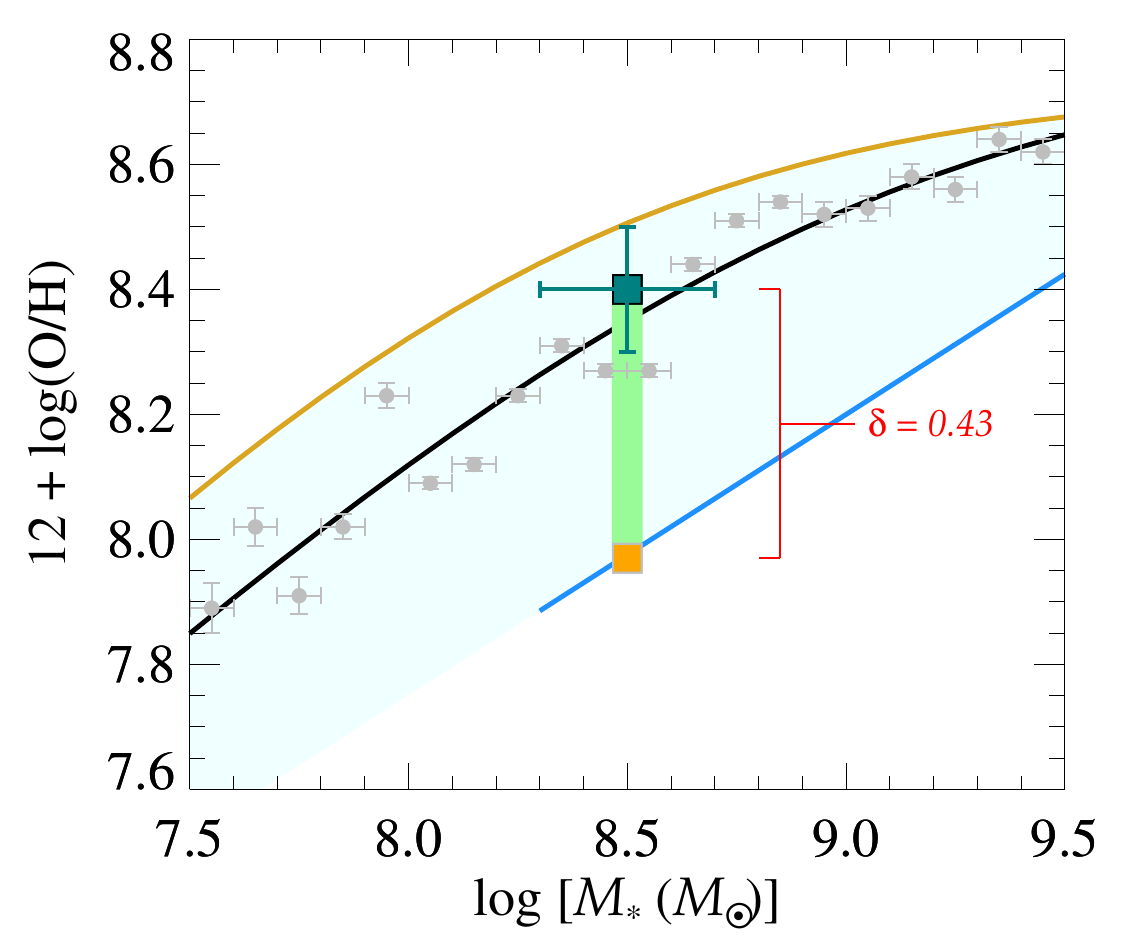}
\caption{Comparison of the metallicity \Zemm\ and stellar mass of \thisgal
  , to the stellar mass-metallicity-SFR (\mzs ) relationship for 200,000
  SDSS galaxies studied by \citet{andrews13}. The black line represents the
  \mzs\ relation found without considering the SFR of the sample galaxies;
  the grey circles show the data used to produce the black line and are
  included to indicate the dispersion in the data.  The yellow and blue
  lines show the resulting \mzs\ relations when the galaxies are binned into
  a sample with low SFRs, $-1.0\: \leq\: \log$(SFR)$\:< -0.5$, and 
  with high SFRs, $0.5\:\leq\: \log$(SFR)$\:< 1.0$, respectively. The green
  square shows the position of \Zemm\ for \thisgal , while the orange
  square shows the value of \Zabs. The difference between the two, the
  value of $\delta$ discussed in this paper, is of the same magnitude as
  the \mzs\ purports to measure for different SFRs. This suggests
  that the baseline metallicity of a galaxy should be that of the neutral
  ISM, as it may be significantly different than the emission line
  metallicities that define the \mzs\ relationship. \label{fig_MZR}}
\end{center}
\end{figure}

In addition, our results may have more general consequences for the
calibration of the well known mass-metallicity relation, MZR, and the
mass-metallicity-SFR correlation, \mzs\ \citep[][and
refs.~therein]{hirschauer18}. Galaxy evolution models attempt to reproduce
the slope, shape and scatter within these relationships (despite the
difficulties in the absolute calibration of O/H discussed in
\S\ref{sect_h2}).  The position of \thisgal\ on the MZR diagram is shown in
Figure~\ref{fig_MZR}, which in this particular case uses the results
constructed by \citet{andrews13}. Both \Zabs\ and \Zemm\ found in \thisgal\
are plotted.  The difference between the two is of a magnitude similar to
the difference between the full range of correlations that exist for
different ranges of the SFRs, from $-1.0\:\leq\:\log$(SFR)$\:\leq
1.0$. This suggests that positive values of $\delta$ of the magnitude
discussed in this paper could be an important factor in the calibration of
the \mzs\ relation, and that models and numerical simulations that attempt
to explain this important correlation ought to start with a lower baseline
metallicity for galaxies than those that currently employ \Zemm .

As yet, we do not know if positive values of $\delta$ exist for much more
massive galaxies. Some absorption line measurements of gas outside of the
local ISM (along sightlines toward background QSOs) support the idea of
infalling low-metallicity gas \citep{ribaudo11} and even show evidence of
the mixing of outflowing and inflowing material in their CGM
\citep{frye19}.  At distances of many tens of kpc there appears to be
significant offsets in \Zemm\ and the metallicity of the CGM
\citep{kacprzak19}.  Within higher-mass late-type galaxies, star formation
processes are clearly modified by factors not experienced by dwarf
galaxies, such as the density waves which define the former's spiral
structure. Strong negative metallicity gradients clearly demonstrate that
the metallicity of ionized gas (\Zemm) changes with galactocentric radius,
and there is some evidence that there are local variations imprinted on the
radial gradients that may be due to inflows from the IGM
\citep{howk18,hwang19} or mixing-induced dilution of the metals by the
spiral density waves passing through the disk \citep{ho17}. These ideas can
eventually be tested by measuring \Zabs\ towards multiple \ion{H}{2} sites
in large local disk-galaxies and by mapping variations in both \Zabs\ and
$\delta$ over kpc and sub-kpc scale lengths.

\begin{deluxetable*}{clccccccccccc}
\tabletypesize{\footnotesize}
\tablecolumns{13} 
\tablecaption{Emission and Absorption Line Metallicities from Individual Galaxies\label{tab_collection} }
\tablehead{
\colhead{}
& \colhead{}
& \colhead{H~II}
& \colhead{H~II}
& \multicolumn{4}{c}{H~II metallicities}
& \colhead{H~I}
& \colhead{H~I}
& \multicolumn{3}{c}{H~I metallicities}
\\
\cline{5-8}\cline{11-13}
\colhead{ID}
& \colhead{Galaxy}
& \colhead{Ref.}
& \colhead{Aper.}
& \colhead{Ar3O3}
& \colhead{N2}
& \colhead{O3N2}
& \colhead{KK04}
& \colhead{Ref.}
& \colhead{Aper.}
& \colhead{\Zabs (O)}
& \colhead{\Zabs (S)}
& \colhead{\Zabs (P)}
}
\colnumbers
\startdata
\input{metals-latex-2-Table}
\hline
\enddata
\tablecomments{Columns --- (1): These numbers are used to
  identify galaxies in Fig.~\ref{fig_delta}; (2): galaxy name; (3):
  reference to the work used herein to measure \ion{H}{2} emission line
  ratios (see reference list below);
  (4): name of the aperture used by the authors in column 3, quoted
  verbatim if available, or, approximate size of aperture used to obtain
  emission line data; 
(5$-$8): values of 12+log(O/H)
  metallicities derived from the  emission line ratios. Values of O3N2 are
  included for completeness, even though it has already been established that many
  of the listed galaxies have such low metallicities that the O3N2
  parameter is invalid in these cases;
 (9): reference to the work used to measure absorption lines (see below);
(10): instrument
  aperture used to measure absorption lines. All COS observations used the
  $2\farcs5$ diameter Primary Science Aperture (PSA); the designations ``LWRS'' and ``MDRS'' refer to the
  30x30$\arcsec$ and 4.0x20$\arcsec$ apertures of the FUSE satellite, respectively. The {\it Space Telescope
  Imaging Spectrograph} (STIS) observations of HS~1543+593 were made with
the 52x0.1$\arcsec$ aperture; (11$-$13): metallicities \Zabs\ derived from
absorption lines assuming solar abundances.
}
\tablenotetext{a} {For NGC~3690, there may be a mis-match in the position of
  the absorption and emission line apertures: the position of the SDSS
  fiber used to measure the emission lines is recorded as being $\approx 13\arcsec$ from the
  UV source used by \citet{james14} to measure absorption lines. For
  NGC~5253, the position of  ``NGC~5253$-$2'' listed in \citet{james14} does not correspond to the
  position of ``Aperture 2'' (or ``UV4'') labelled by \citet{calzetti97}, and so
  is not included in this list.
}
\tablerefs{\input{collected_refs}}
\end{deluxetable*}

\bigskip
\section{Summary}

\bigskip
We summarize the results of this paper as follows:

\begin{enumerate} 

\item We have used COS to observe the QSO \qso\ behind the galaxy \thisgal
  . The galaxy has a systemic redshift of $cz=1577$~\kms, and has a
  luminosity, type, and \ion{H}{1} gas mass similar to the SMC. Like the
  SMC, \thisgal\ is part of a group of galaxies, although it is further
  away from its more massive host than the SMC is from the Milky Way.  The
  background QSO was selected specifically because of the identification of
  its sightline through the dwarf galaxy, and the \ion{H}{1} absorption
  detected does not represent an
  unbiased probe of $N$(\ion{H}{1}) in the local universe.  We have used
  the emission lines in an SDSS spectrum of the central \ion{H}{2} region
  to derive a metallicity of \Zemm$\,= -0.37\pm0.10$ in the ionized gas
  using the S23 strong-line ratio, although a selection of other methods
  produce metallicities that agree well with this value.

\item The COS spectrum shows that at a galactocentric radius of
  $\simeq 1$~kpc, \thisgal\ is a DLA absorber, with
  $\log N$(\ion{H}{1})$\:=20.89 \, [+0.12, -0.21]$ at a redshift of
  $cz = 1581$~\kms . Metal lines are detected from the DLA, and the
  analysis of a set of weak lines suggests that the metallicity of the
  neutral gas (assuming no depletion from dust and no need of any
  ionization corrections) is \Zabs$\: = -0.80\pm{0.24}$, which is lower
  than that seen in the ionized \ion{H}{2} region gas by a factor of
  $\sim 3$.

\item The difference in the metallicity seen in absorption and that in
  emission, $\delta$, is higher ($\approx 0.4$ dex) than the value found
  along a previous QSO sightline that intercepted the dwarf galaxy
  SBS~1543+5921, although their errors overlap. \thisgal\ has a value of
  $\delta$ similar to those measured towards bright \ion{H}{2} regions
  within a set of dwarf star-forming galaxies.  Although the errors towards
  individual sightlines are often large, we confirm that collectively, a
  small offset of $\delta\:\approx\: 0.4-0.5$ dex persists in the data.

\item If the evolution of \thisgal\ is such that the metals throughout the
  galaxy have been well mixed on kpc scales, then the simplest model to
  explain the fact that \Zabs$\:<\:$\Zemm\ is that low metallicity gas from
  the IGM has flowed into the galaxy and diluted the gas in the ISM,
  leading to a low value of \Zabs\ along the line of sight to the QSO. This
  model is consistent with the detection of \lya\ emission in the core of
  the damped \lya\ profile at a velocity offset from the galaxy's systemic
  velocity.

\end{enumerate}

\thisgal\ remains a largely unstudied galaxy, and some obvious additional
observations would help elucidate the origin of the DLA better. High
spatial resolution 21~cm maps of the distribution of high $N$(\ion{H}{1})
would show how strongly the galaxy is interacting with its environment,
through, e.g., its morphological asymmetry, the presence of tidal features,
additional \ion{H}{1} companions closer than UGC~5287, etc.  Integral field
spectroscopy of the emission lines across the galaxy is now possible with
modern instruments, and would test whether \Zemm\ changes between the
central \ion{H}{2} region and the QSO sightlines, or whether there exists
dramatic discontinuities in \Zemm\ as a result of strong inflows from the IGM.
Finally, ground-based, high S/N echelle observations of \ion{Ti}{2} in the
spectrum of \qso\ would better constrain the absorption line model measured
in the UV data and could provide an additional estimate of the metallicity
and dust depletion of the neutral gas along the sightline.

\bigskip
\acknowledgments

The work required to identify QSOs behind galaxies was supported by a NASA
{\it Long Term Space Astrophysics} (LTSA) grant NNG05GE26G.  Funding for
the reduction of the COS spectra obtained from the GO program was provided
by NASA grant number 12486 from the Space Telescope Science Institute
(STScI), which is operated by the Association of Universities for Research
in Astronomy, Inc. (AURA), under NASA contract NAS5-26555.  Funding for
{\it Sloan Digital Sky Survey} SDSS-III has been provided by the Alfred
P.~Sloan Foundation, the Participating Institutions, the National Science
Foundation (NSF), and the U.S. Department of Energy Office of Science.
CALCOS and PyRAF are products of the STScI, which is operated by AURA for
NASA. IRAF, on which parts of PyRAF are based, is distributed by the
National Optical Astronomy Observatory, which is operated by AURA under a
cooperative agreement with the NSF. This publication makes use of data
products from the WISE satellite, which is a joint project of the
University of California, Los Angeles, and the Jet Propulsion
Laboratory/California Institute of Technology, funded by NASA.  The
STARLIGHT project is supported by the Brazilian agencies CNPq, CAPES and
FAPESP and by the France-Brazil CAPES/Cofecub program. This research has
made use of the NASA/IPAC Extragalactic Database (NED), which is operated
by the Jet Propulsion Laboratory, California Institute of Technology, under
contract with NASA.

\facilities{HST (COS), WISE, GALEX, ARC (SPIcam, DIS), SDSS}


\bibliographystyle{aasjournal}
\bibliography{ms9}

\end{document}

%% file: metals-latex-2-Table.tex
1  &SBS~0335$-$052 E &1  &Nos.~4+5 FORS &7.15 &7.61 &7.91 &7.65 &2   &COS         &$-3.69\, (\pm 0.07)$      &$-2.00\, (\pm 0.07)$       &$-2.27\, (\pm 0.16)$\\
   &                 &1  &Nos.~4+5 UVES &6.98 &7.59 &7.89 &7.72 &    &            &&&\\
   &                 &3  &$3\farcs5$    &7.00 &7.54 &7.87 &7.68 &    &            &&&\\
2  &I~Zw18~NW        &4  &$3\arcsec$    &7.47 &7.63 &7.95 &\ldots    &2,5 &COS         &$-1.69\, (\pm 0.06)$      &$-1.83\, (\pm 0.08)$       &$-1.92\, (\pm 0.31)$\\
   &                 &6  &NW-knot       &7.55 &7.58 &7.93 &7.63 &    &            &&&\\
   &                 &3  &$3\farcs5$    &7.53 &7.47 &7.88 &\ldots    &    &            &&&\\
   &                 &7  &P0-B0         &7.74 &7.58 &7.94 &7.60 &    &            &&&\\
   &                 &7  &P0-C0         &7.53 &7.59 &7.93 &7.56 &    &            &&&\\
3  &I~Zw~36          &4  &$3\arcsec$    &7.72 &7.77 &7.91 &\ldots    &8   &     LWRS   &$-1.26\, (^{+1.85}_{-1.02})$ & \ldots                         &$-1.84\, (^{+0.30}_{-0.38})$\\
   &                 &3  &$3\farcs5$    &7.75 &\ldots    &\ldots    &8.18 &    &            &&&\\
4  &SBS 1415+437     &9  &n/a           &7.84 &7.87 &8.01 &8.04 &2   &COS         &$-2.98\, (\pm 0.11)$      &$-1.39\, (\pm 0.05)$       &$-1.20\, (\pm 0.09)$\\
   &                 &10 &e1            &7.83 &7.86 &8.00 &7.99 &    &            &&&\\
   &                 &11 &$5\arcsec$    &7.78 &7.88 &8.01 &8.02 &    &            &&&\\
   &                 &11 &$0\farcs6$    &7.75 &7.83 &7.99 &7.99 &    &            &&&\\
5  &NGC 1705         &12 &s7-1          &8.17 &8.10 &8.13 &8.22 &13  &     LWRS   &$-1.33\, (\pm 0.22)$      &\ldots                          &\ldots\\
   &                 &12 &s7-2          &8.10 &8.08 &8.10 &8.27 &    &            &&&\\
   &                 &12 &s8-1          &8.05 &8.04 &8.07 &8.19 &    &            &&&\\
   &                 &12 &s8-2          &8.06 &8.07 &8.09 &8.20 &    &            &&&\\
   &                 &12 &s8-3          &8.08 &8.12 &8.12 &8.26 &    &            &&&\\
   &                 &12 &s8-4          &8.13 &8.19 &8.17 &8.28 &    &            &&&\\
6  &Pox 36           &14 &$5\farcs5$    &8.13 &8.08 &8.10 &8.35 &15  &     LWRS   &$-1.52\, (\pm 0.13)$      &\ldots                         &$-0.88\, (\pm 0.18)$\\
   &                 &16 &$4\arcsec$    &\ldots    &8.10 &8.10 &8.37 &    &            &&&\\
7  &NGC 4670         &4& $3\arcsec$     &8.26 &8.19 &8.17 &\ldots    &2   &COS         &$-2.57\, (\pm 0.10)$      &$-0.73\, (\pm 0.09)$       &$-0.79\, (\pm 0.10)$\\
   &                 &17 &Region 1      &\ldots    &8.14 &8.12 &8.15 &    &            &&&\\
8  &NGC 5253$-$1\tablenotemark{a}     &18 &UV2           &8.31 &8.23 &8.19 &8.36 &2   &COS     &$-2.52\, (\pm 0.15)$      &$-1.02\, (\pm 0.05)$       &$-1.00\, (\pm 0.09)$\\
9  &NGC 604          &19 &$6\arcsec$    &8.36 &8.27 &8.25 &7.88 &20  &Various     &$-1.21\, (\pm 0.30)$      &\ldots                          &$-0.59\, (\pm 0.32)$\\
   &                 &21 &n/a           &8.45 &8.33 &8.27 &8.08 &    &            &&&\\
10 &NGC 3690\tablenotemark{a}         &4  &$3\arcsec$    &8.42 &8.56 &8.44 &\ldots    &2   &COS      &$-1.90\, (\pm 0.06)$      &$-0.46\, (\pm 0.06)$       &\ldots\\
11 &NGC 4214         &22 &n/a           &\ldots   &8.25 &8.20 &8.45 &2   &COS         &$-2.63\, (\pm 0.06)$      &$-0.82\, (\pm 0.06)$       &$-0.90\, (\pm 0.06)$\\
12 &NGC 4449         &23 &main reg.     &\ldots    &8.37 &8.29 &8.33 &2   &COS         &$-2.36\, (\pm 0.06)$      &$-0.79\, (\pm 0.10)$       &$-0.57\, (\pm 0.10)$\\
   &                 &22 &n/a           &\ldots    &8.31 &8.27 &8.31 &    &            &&&\\
13 &NGC 625          &24 &No.~5         &\ldots    &8.11 &8.10 &8.26 &25  &     MDRS   &$-1.46\, (\pm 0.33)$      &\ldots                          &$-0.50\, (\pm 0.34)$\\
   &                 &24 &No.~9         &\ldots    &8.21 &8.19 &8.19 &    &            &&&\\
   &                 &24 &No.~18        &\ldots    &8.27 &8.23 &7.69 &    &            &&&\\
   \multicolumn{13}{c}{QSOs behind galaxies}\\
   \hline
14 &SBS 1543+5921 +   &26 &n/a           &\ldots    &8.18 &8.15 &8.34 &27  &STIS        &\ldots                         &$-0.49\, (\pm 0.08)$       & \ldots\\
   &HS 1543+593     &   &              &     &     &     &     &    &            &&&\\
15 &UGC 5282 +       &4&$2\arcsec$      &8.43 &8.26 &8.27 &8.36 &4   &COS         &\ldots                        &$-0.82\, (^{+0.39}_{- 0.26})$ &$-0.78\, (^{+0.30}_{-0.32})$\\
   &Q0951+3307       & &                &     &     &     &     &    &            &&&\\

%% file: collected_refs.tex
(1) --- \citet{izotov09};
(2) --- \citet{james14};
(3) --- \citet{izotov97};
(4) --- this paper;
(5) --- \citet{lebout13}
(6) --- \citet{kehrig16};
(7) --- \citet{vilchez98};
(8) --- \citet{lebout04};
(9) --- \citet{melbourne04};
(10) --- \citet{guseva03};
(11) --- \citet{thuan99};
(12) --- \citet{annibali15};
(13) --- \citet{heckman01};
(14) --- \citet{izotov04};
(15) --- \citet{lebout09};
(16) --- \citet{kunth83};
(17) --- \citet{kumari18};
(18) --- \citet{lopez-sanchez07};
(19) --- \citet{esteban09};
(20) --- \citet{lebout06};
(21) --- \citet{vilchez88};
(22) --- \citet{kobulnicky99};
(23) --- \citet{kumari17};
(24) --- \citet{skillman03};
(25) --- \citet{cannon05};
(26) --- \citet{regina05};
(27) --- \citet{sbs1543_2}.
In this paper (Ref ``4'' in column 3), the following SDSS spectra were
used:
I~Zw~18 NW --- 0555-52266-0558;
I~Zw~36 --- 1453-53084-0322;
NGC~4670 --- 2238-54205-0222;
NGC~3690 --- 0952-52409-0247.